\documentclass[fleqn,usenatbib,useAMS]{mnras}
\usepackage{hyperref}
\hypersetup{draft}
\usepackage{graphicx}	
\usepackage{amsmath}	
\usepackage{amssymb}	
\usepackage{multicol}        
\usepackage{bm}		

\usepackage[T1]{fontenc}
\usepackage{ae,aecompl}
\usepackage{txfonts}
\usepackage{gensymb}
\usepackage{url}
\usepackage{txfonts}

\title[HERBS II: Light, iron-peak and neutron-capture elements]{HERBS II: Detailed chemical compositions of Galactic bulge stars}
\author[L. Duong et al.]{L. Duong,$^{1}$ M. Asplund,$^{1,2}$\thanks{Contact email: \href{mailto:martin.asplund@anu.edu.au}{martin.asplund@anu.edu.au}} D. M. Nataf,$^{3}$ K. C. Freeman$^{1}$ and M. Ness$^{4,5}$ 
	\\
	$^{1}$Research School of Astronomy \& Astrophysics, Australian National University, ACT 2611, Australia\\
	$^{2}$ARC Centre of Excellence for All Sky Astrophysics in 3 Dimensions (ASTRO 3D)\\
	$^{3}$Center for Astrophysical Sciences and Department of Physics and Astronomy, The Johns Hopkins University, Baltimore, MD 21218, USA\\
	$^{4}$Department of Astronomy, Columbia University, Pupin Physics Laboratories, New York, NY 10027, USA\\
	$^{5}$Center for Computational Astrophysics, Flatiron Institute, 162 Fifth Avenue, New York, NY 10010, USA\\
}

\date{Accepted XXX. Received YYY; in original form ZZZ}

\pubyear{2018}

\begin{document}
\label{firstpage}
\pagerange{\pageref{firstpage}--\pageref{lastpage}}
\maketitle

\begin{abstract}
	
\noindent This work explores the detailed chemistry of the Milky Way bulge using the HERMES spectrograph on the Anglo-Australian Telescope. Here we present the abundance ratios of 13 elements for 832 red giant branch and clump stars along the minor bulge axis at latitudes $b=-10^{\circ}, -7.5$ and $-5^{\circ}$. Our results show that none of the abundance ratios vary significantly with latitude. We also observe {disk-like} [Na/Fe] abundance ratios, which indicates the bulge does not contain helium-enhanced populations as observed in some globular clusters. Helium enhancement is therefore not the likely explanation for the double red-clump observed in the bulge. We confirm that bulge stars mostly follow abundance trends observed in the disk. However, this similarity is not confirmed across for all elements and metallicity regimes. The more metal-poor bulge population at [Fe/H] $\lesssim -0.8$ is enhanced in the elements associated with core collapse supernovae (SNeII). In addition, the [La/Eu] abundance ratio suggests higher $r$-process contribution, and likely higher star formation in the bulge compared to the disk. This highlights the complex evolution in the bulge, which should be investigated further, both in terms of modelling; and with additional observations of the inner Galaxy.
\end{abstract}



\section{Introduction}
The bulge region of the Milky Way is a complex system that has been the focus of many recent Galactic studies (see \citealt{Nataf2016,Barbuy2018} and references therein). Once thought to be an exclusively old, classical bulge, much evidence has emerged suggesting it is a pseudo-bulge, formed via disk instability. Infrared images of the Galactic bulge show its X-shaped morphology, which is supported by the double clump seen in photometric studies and kinematic studies confirmed that it rotates cylindrically~\citep{Dwek1995,McWilliam2010,Kunder2012,Ness2013a,Nataf2015,Ness2016b}. Furthermore, the chemistry of the bulge, in particular, the abundances of alpha elements\footnote{Typically, alpha elements are O, Mg, Si, Ca, Ti, see e.g.~\cite{Burbidge1957}.} largely follow the high-$\alpha$ disk trend~\citep{Melendez2008,Alves-Brito2010, Gonzalez2011,Johnson2014,Bensby2017,Rojas-Arriagada2017,Jonsson2017}. 

There are, however, hints in the data indicating that bulge stars may have experienced a slightly different chemical evolution to the high-$\alpha$ disk. Based on their high-resolution data of bulge microlensed dwarfs and turn-off stars,~\cite{Bensby2013,Bensby2017} found evidence that the `knee' of the bulge, or the metallicity at which [$\alpha$/Fe] begins to decline, is more metal-poor compared to the local high-$\alpha$ disk. Because the alpha elements are thought to be produced in massive stars ($M \geq 8 \ M_\odot$) with short lifetimes, this suggests faster chemical enrichment of the bulge region~\cite[e.g.][]{Tinsley1979,Matteucci1990}. By comparing the alpha abundances of stellar samples from many different studies,~\cite{McWilliam2016} found offsets between the bulge and the high-$\alpha$ disk, such that the bulge is more enhanced in the alpha elements. However, this conclusion is still under debate: the difference between the high-$\alpha$ disk alpha knee and bulge is small, and may not be significant given measurement errors~\citep{Bensby2017,Rojas-Arriagada2017}. Furthermore, offsets found between independent bulge and disk studies may well be due to the variations in linelists, atomic data and analysis techniques~\citep{Haywood2018}. 

Abundance measurements for element groups other than the alpha elements are also crucial to our understanding of chemical evolution in the Galactic bulge. The light elements Na and Al have unique anti-correlation signatures {with respect to O and Mg, respectively, }which can inform the origin of bulge stars (\citealt{Gratton2012,Bastian2017}, and references therein). It has also been demonstrated that, for globular clusters, relatively high [Na/Fe] indicates a helium-enhanced stellar population~\citep[e.g.,][]{Carretta2010,Dupree2011}. In addition, the bulge [Al/Fe] vs [Fe/H] trend is similar to the alpha elements, which indicate that aluminium is produced in massive stars and will therefore provide formation time-scale constraints. 

The iron-peak elements, although part of the same group, typically do not share the same production site and for some elements there are large discrepancies between predicted and observed abundance ratios~\citep[e.g.][]{Romano2010}. The abundance ratios of iron-peak elements also display different trends with metallicity, not necessarily tracking iron. Scandium is thought to be produced mostly in massive stars~\citep{Woosley1995}, and shares a similar behaviour to the alpha elements. Cu is also thought to be produced in massive stars but follows an interesting, non-alpha trend, both in the Galactic disk and bulge~\citep{Woosley1995,Reddy2006,Romano2007,Johnson2014}. However, Cu abundance measurements for bulge field stars are rare in the literature, with results from \cite{Johnson2014} being the only data available. On the other hand, Cr, Mn, Co and Ni are produced via silicon burning processes~\citep{Woosley1995}, but their correlation with metallicity can be very different~\citep[e.g.][]{Johnson2014,Bensby2017}. 

Neutron-capture elements are produced via the $r$- (rapid) or $s$- (slow) process. The $r$-process is thought to trace rapid enrichment timescales, as one of the possible formation sites is core-collapsed supernovae, or SNeII (see e.g. \citealt{Woosley1994}; \citealt{Sneden2008} and references therein). The merging of neutron stars is now more favoured as the likely production site of the $r$-process~\citep[e.g.,][]{Thielemann2017,Cote2018}. The $s$-process occurs in low- and intermediate-mass AGB stars, with lifetimes up to several Gyrs~\citep[e.g.,][]{Karakas2014}. Neutron-capture elements are produced by a combination of both $s$- and $r$- processes. For certain elements, the contribution from one process is much greater the other, thus they are referred to as `$r$-process' or `$s$-process' elements. Commonly studied neutron-capture elements are Ba and Eu, which {in the Sun} have $\approx$85\% and $\approx$5\% $s$-process contribution, respectively~\citep{Sneden2008,Bisterzo2014}.

There are a few gaps in the literature regarding the abundance trends of bulge stars. One is that few measurements exist for metal-poor stars between $-2 \leq$ [Fe/H] $\leq -1$, while an extensive study of extremely metal-poor stars have been conducted by~\citep{Howes2015,Howes2016}. Additionally, the number of stars with abundance measurements for elements other than alpha is still relatively small. This is mainly due to difficulties with observing faint, reddened bulge targets at high resolution. As outlined above, having abundances of other elements, such as the light and neutron-capture elements will be informative for the formation/evolution timescale of the Galactic bulge. The HERMES Bulge Survey (HERBS) was designed to address some of these gaps. As detailed in \cite{Duong2019} (hereafter Paper I), we made use of the extensive wavelength coverage of the HERMES spectrograph, which can provide abundances for up to 28 elements. The high multiplexity of the 2dF/AAT system ($\approx$350 science objects observed in a single exposure) allows for longer integration times and thus adequate signal to noise ratio (S/N) for abundance determination. One advantage of HERBS is its compatibility with abundance measurements of the GALAH survey, which targeted disk and halo stars. By using the same atomic data and analysis method, we are able to eliminate many systematic offsets that may affect the bulge-disk comparison. 

In Paper I, we have presented the HERBS survey and detailed the data analysis. There we also provided the stellar parameters and abundance ratios of five alpha elements (O, Mg, Si, Ca, Ti) for 832 bulge giants. In this paper we present and discuss the results of 13 more elements from different nucleosynthesis channels: the light elements Na and Al; iron-peak elements Ni, Mn, Cu, Cr, Co, Sc; neutron capture elements La, Nd, Eu, Ce and Zr.

\section{Data description and analysis}

The details of data selection, observation, reduction and analysis for the stellar sample presented here have been discussed in Paper I, which the reader is referred to for more information. Briefly, the majority of the sample are red giant and red clump stars with confirmed bulge membership from the ARGOS survey~\citep{Freeman2013,Ness2013}. The 2MASS $K_s$ magnitude range of ARGOS stars is 11--14. Our sample also includes 15 bulge stars from the EMBLA survey~\citep{Howes2016}, which were added to help increase the number of metal-poor stars. We were able to probe the full metallicity range of the bulge, from [Fe/H] $\approx 0.5$ to [Fe/H] $\approx -2$. The minor axis fields $(\ell,b)=(0,-10)$; $(0,-7.5)$ and $(0,-5)$ were observed with the HERMES spectrograph, at resolving power $\mathcal{R} \approx 28 000$. The four wavelength intervals covered by HERMES are 4713--4903~\AA~(blue CCD); 5648--5873~\AA~(green CCD); 6478--6737~\AA~(red CCD) and 7585--7887~\AA~(IR CCD). 

All spectra were reduced with the standard HERMES reduction software 2dfdr v6.46\footnote{\url{www.aao.gov.au/science/software/2dfdr}}. Additional data processing (barycentric correction, telluric correction, co-adding) were done with custom \textsc{python} scripts. We used the GUESS code, which is also used by the GALAH survey, to determine radial velocities and estimate initial stellar parameters~\citep{Kos2017}. The final stellar parameters were obtained with the spectral synthesis software \emph{Spectroscopy Made Easy}~\citep{Valenti1996,Piskunov2017} using 1D LTE \textsc{marcs} model atmospheres~\citep{Gustafsson2008}. Starting from initial estimates, SME solves for best-fit stellar parameters by optimising $\chi^2$ through an iterative process. {During the stellar parameter determination stage, we directly implement non-LTE departure coefficients from \cite{Amarsi2016b} for \ion{Fe}{i} lines}. All lines and atomic data used to derive stellar parameters are the same as GALAH Survey DR2~\citep{Buder2018}. In total there are 313 stars in field $(0,-10)$, 313 stars in $(0,-7.5)$ and 204 stars in field $(0,-5)$ with reliable stellar parameters. {The median signal-to-noise ratios for each field are given in Table \ref{table:snr}.

\begin{table}
	\caption{The estimated $V$-magnitude and median signal-to-noise ratio of each bulge field. The IR arm is not shown as it has similar SNR to the red arm. For HERMES, one resolution element is equivalent to approximately four pixels.}
	\label{table:snr}
	\begin{tabular}{llllll}
		\hline 
		Field & RC   & Exp time & SNR$_\mathrm{B}$ & SNR$_\mathrm{G}$& SNR$_\mathrm{R}$ \\
		($\ell,b$)  &$V$\textsubscript{mag} & (hours) & (pixel$^{-1}$) &  (pixel$^{-1}$) & (pixel$^{-1}$)  \\
		\hline
		(0, $-5$) & 17.4 & 17 &  20 &  34 &  46\\
		(0,$-7.5$) & 16.3 & 10 & 32 &  51 &  65\\
		(0, $-10$) & 16.0 & 08 & 30 &  40 &  53\\
		\hline
	\end{tabular}
\end{table}
}
Elemental abundance ratios were determined by SME after stellar parameters have been established, also using $\chi^2$-optimisation. Pre-determined wavelength regions covering each line (line masks) are given for each element. During the optimisation stage, SME computes a model spectrum based on the stellar parameters and atomic data given, and de-selects blended wavelength points within line masks. {The full linelist also contains molecules such as CN and TiO, and where applicable these lines are included the abundance synthesis.}

We have used the same atomic data as the GALAH survey, and where possible, the exact same lines. However, we optimised individual lines then computed the weighted average, whereas GALAH optimised all lines simultaneously for each element. In our analysis, we discovered that certain lines show problematic behaviour, such as spuriously high abundances at the metal-poor and metal-rich regimes owing either to unknown blends or possible non-LTE effects. To provide as accurate results as possible, we excluded these lines from the final abundance ratios. While every effort was made to be consistent with the GALAH survey, some of the linelist changes we implemented could not be adopted in time for the GALAH Data Release 2 (DR2). As the result, GALAH would have used more lines than we did here for some elements. {We discuss further the differences between our abundances and that of GALAH in Section \ref{sec4} and Appendix \ref{herbs-galah-comp}. The full list of lines used in this work is given in Appendix \ref{a1}}.

{Some of the lines selected for elements Cu, La and Eu show hyperfine splitting. In addition, Cu and Eu are sensitive to the $^{63/65}$Cu and $^{151/153}$Eu isotope ratios. The hyperfine broadening components are included in the synthesis for \ion{La}{ii} at 4804.069 \AA, which is the only \ion{La}{ii} line with hyperfine broadening data \citep{Lawler2001a}. Both the hyperfine broadening and isotope ratios are included in the line synthesis of \ion{Cu}{i} and \ion{Eu}{ii} \citep{Heiter2015a,Lawler2001b}. The $\log(gf)$ values in Table \ref{table:atom} for these three lines reflect the total value of all hyperfine broadening and isotope components.}

To compute the final abundance ratios, we averaged the abundance ratios of individual lines, normalised to solar values and weighted by $chi^2$ uncertainties. We do not include abundances that are flagged as upper limits in the weighted average. For solar normalisation, we used abundance ratios from a HERMES twilight spectrum, which was reduced and analysed in the same manner as a typical star (see Paper I for details). This ensures systematic errors, such as uncertain $\log \left(gf\right)$ values are mostly removed. For Na and Al, we took into account departures from LTE using the computations by~\cite{Lind2011} and~\cite{Nordlander2017b}, respectively. For all other elements, abundance ratios were computed assuming LTE. {Finally, we adopted the SME $\chi^2$ synthesis errors as our uncertainties. We acknowledge that these uncertainties are underestimated; however in this way we remain consistent with the GALAH survey, which aids the comparison of the two surveys.}

Although HERMES wavelength coverage in principle allows 28 elements to be measured, we have had to omit a number of elements for various reasons, as outlined below:
\begin{itemize}
\item The light element Li can only be measured with lines that are too weak for reliable abundance determination, at least for the RGB stars in this study.
\item The light element K (7699 \AA) shows strong temperature dependency, and large scatter likely caused by non-LTE effects and interstellar absorption that are unaccounted for. 
\item The iron-peak elements V and Zn only have lines in the blue arm of HERMES. Since this arm has the lowest S/N and is most affected by background blends, the abundance ratios show large scatter and/or unexpected trends.  
\item  The abundance ratios of neutron-capture elements Rb (7800 \AA), Sr (6550 \AA), and Sm (4854.5 \AA) can only be determined with a single, weak line at HERMES resolving power. In particular the Rb line is blended by a neighbouring \ion{Si}{i} line. Upon inspection of spectral syntheses, it is evident that the measurements are not reliable, especially not at low S/N. 
\item Yttrium lines are blended or fall in regions where continuum determination is very difficult. The scatter in [Y/Fe] is $\approx$ 1 dex. We are not confident that these results are reliable. 
\item Similar to GALAH survey, our results for neutron-capture element Ba show more than 1 dex scatter for both the 5854 and 6497 \AA~lines at all metallicities. To date we have not found the reason for this behaviour, but it is being investigated.
\end{itemize}

For most elements, the abundance ratios reported below do not show effective temperature dependence. Only the elements Mn and Co show significant trends with $T_\mathrm{eff}$, but they are known to be affected by non-LTE effects~\citep{Bergemann2008, Bergemann2010}. More details can be found in Appendix \ref{a2}. 

\section{Abundance trends}

In this section we discuss the [X/Fe] trends with respect to metallicity. We were able to measure abundances of most elements for [Fe/H] $\geq -1.5$. For each element, we compare our results with recent high-resolution measurements for bulge field stars that span a similar metallicity range. For elements where none, or only one bulge sample is available, abundance ratios for disk stars have been included for reference. The literature samples below all assume LTE in their abundance calculations. 

\subsection{Light elements} 

\begin{itemize}

\item \textbf{Sodium} Our [Na/Fe] trend  is similar to what \cite{Bensby2017} observed for microlensed dwarfs, but with larger scatter (due to lower resolution and S/N). [Na/Fe] values are enhanced low metallicities, decreasing to solar as [Fe/H] increases, and increases to above solar again for [Fe/H] $> 0$. This behaviour indicates that sodium is at least partly produced in massive stars, as [Na/Fe] decreases with metallicity in the sub-solar regime~\citep{Woosley1995}. The results from \cite{Johnson2014} (also for RGB stars) show a different behaviour, where [Na/Fe] is under-abundant at the metal-poor regime. Since \cite{Johnson2014} have shown that the discrepancy is not likely due to the non-LTE corrections from \cite{Lind2011}, the cause may be different lines/atomic data used.   

\item \textbf{Aluminium} The Al abundance trend is basically the same as the alpha elements, in agreement with \cite{Johnson2014} and \cite{Bensby2017}. [Al/Fe] is enhanced by +0.4 dex at low metallicities and decreases with increasing [Fe/H]. This trend has also been observed by other authors for disk stars, and indicates SNeII origin for Al~\citep[e.g.,][]{Reddy2006,Bensby2014}. There is an apparent offset of $\approx$0.1 dex around solar metallicity between our abundance ratios and that of the literature samples. 
\end{itemize}

{While Na and Al show disk-like behaviour in our metallicity regime, \cite{Howes2016} observed particularly low abundance ratios for these two elements for extremely metal-poor bulge stars.}

\begin{figure}
	\centering
	\includegraphics[width=1\columnwidth]{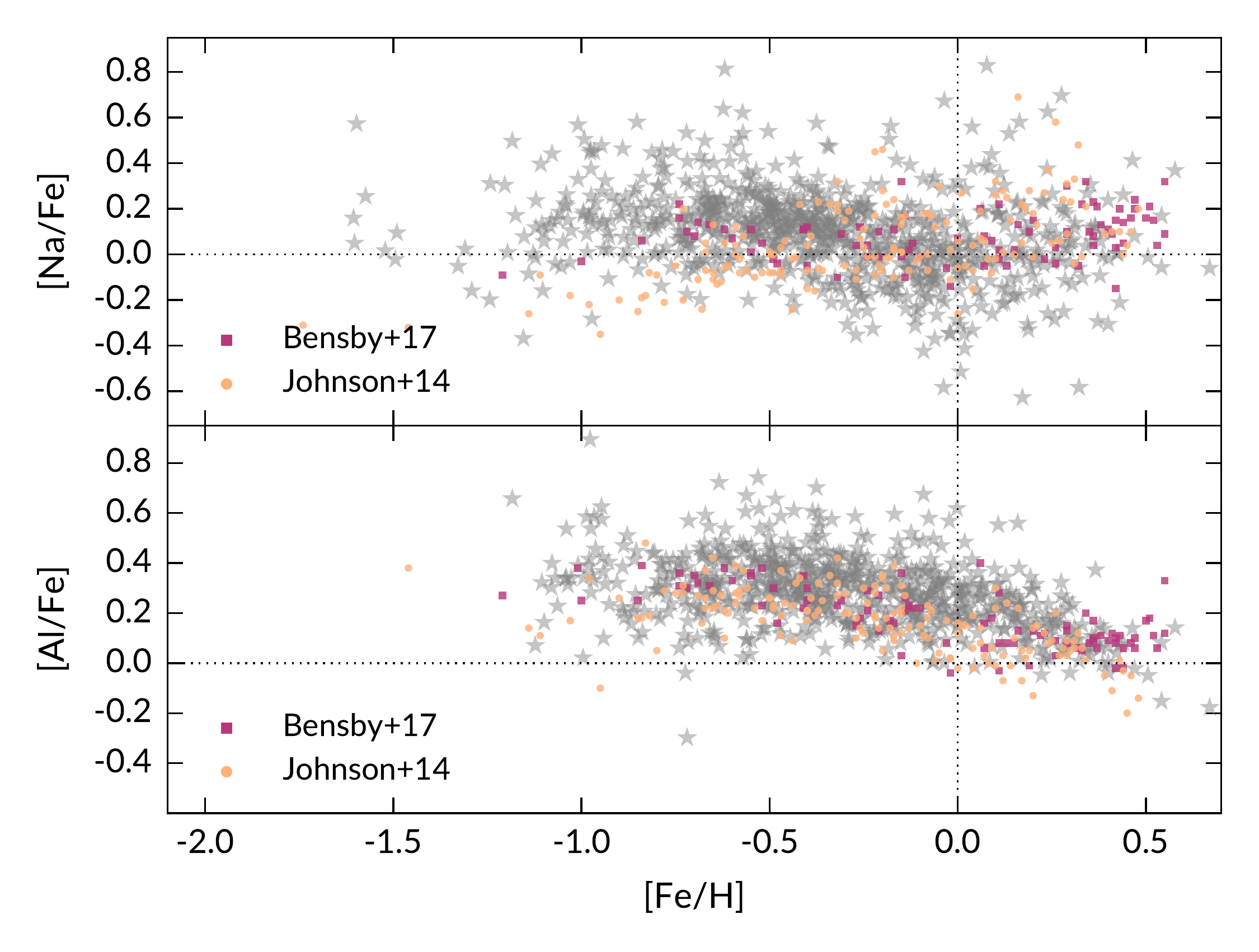}
    \caption{The [X/Fe] vs [Fe/H] trends for the light elements Al and Na from this work (grey stars). Shown for comparison are bulge field stars from \protect\cite{Johnson2014} and \protect\cite{Bensby2017}.}
    \label{fig:lightelems}
\end{figure}

\subsection{Iron-peak elements}

\begin{itemize}
\item \textbf{Scandium} To the best of our knowledge, only our study so far has provided scandium abundance trend for {bulge field stars over a large metallicity range (but see also \citealt{Gratton2006, Casey2015, Howes2016})}. The Sc abundance ratios follow a well defined, alpha-like trend, in good agreement with results from \cite{Battistini2015} for the local disk. The [Sc/Fe] enhancement at low metallicities is not as high as observed in aluminium and the alpha elements (at maximum [Sc/Fe] $\approx$ 0.2 dex), but we do see a decrease in [Sc/Fe] with metallicity, reaching solar values at [Fe/H] $\approx 0$. This behaviour is indicative of scandium production in massive stars, as suggested by~\cite{Woosley1995}. 

\item \textbf{Chromium} [Cr/Fe] ratios show very little scatter, and follows the same trend as observed by \cite{Johnson2014} and \cite{Bensby2017}. [Cr/Fe] remains constant (at solar value) across the entire metallicity range. 

\item \textbf{Manganese} Unfortunately we were not able to measure manganese for {many stars more metal-rich than [Fe/H] = $-0.5$}. All Mn lines are in the blue arm of HERMES, which has the lowest S/N, and are susceptible to blends at higher metallicities. The abundances that we can measure show much larger scatter than those from \cite{Barbuy2013} (bulge giants) and \cite{Battistini2015} (disk dwarfs). However, the general trend is similar: that [Mn/Fe] increases linearly as a function of metallicity, and rises to [Mn/Fe] = 0 at super solar metallicity. Note that \cite{Battistini2015} observed a very different [Mn/Fe] vs [Fe/H] trend when non-LTE corrections are applied. The NLTE manganese abundances become essentially flat, and remains sub-solar at all metallicities. 

\item \textbf{Cobalt} Co is another iron-peak element that shows alpha-like behaviour, as shown by \cite{Battistini2015} for disk main-sequence stars. The alpha-like trend is much less pronounced in our [Co/Fe] values, which tracks iron for [Fe/H] $> -0.5$, and only show a slight increase at lower metallicities. Our results are in good agreement with \cite{Battistini2015}. We do not observe the enhanced [Co/Fe] ratios around solar metallicity as did \cite{Johnson2014}.

\item \textbf{Nickel} Our nickel abundance trend is in agreement with both \cite{Johnson2014} and \cite{Bensby2017}, but with larger scatter. The mean [Ni/Fe] observed here is in agreement with \cite{Johnson2014}, but slightly enhanced compared to \cite{Bensby2017}. The [Ni/Fe] trend is reminiscent of [Na/Fe]: for both elements the abundance ratios are approximately solar between $-0.5<$ [Fe/H] $< 0$ and increase as a function of metallicity at the super-solar regime. However, nickel does not show strong enhancement at the metal-poor regime and the abundance ratios basically remain constant between $-1<$ [Fe/H] $< 0$. 

\item \textbf{Copper} The copper abundance trend in the bulge is not well established; \cite{Johnson2014} was the only study that measured copper abundance ratios for bulge field stars that cover a wide metallicity range. The copper abundance pattern we observe here is similar to that of \cite{Johnson2014}: [Cu/Fe] increases as a function of metallicity for $-1 < $ [Fe/H] $< -0.5$, remains approximately constant between $-0.5 <$ [Fe/H] $< 0$, but appears to increase with metallicity at the super-solar regime. This is consistent with disk [Cu/Fe] measurements from \cite{Reddy2006}. However, \cite{Johnson2014} observe a +0.4 dex enhancement in [Cu/Fe] around $-0.5<$ [Fe/H] $< 0$, which may be due to blended copper lines in their analysis~\citep{McWilliam2016}.

\end{itemize}

\begin{figure*}
	\centering
	\includegraphics[width=1\columnwidth]{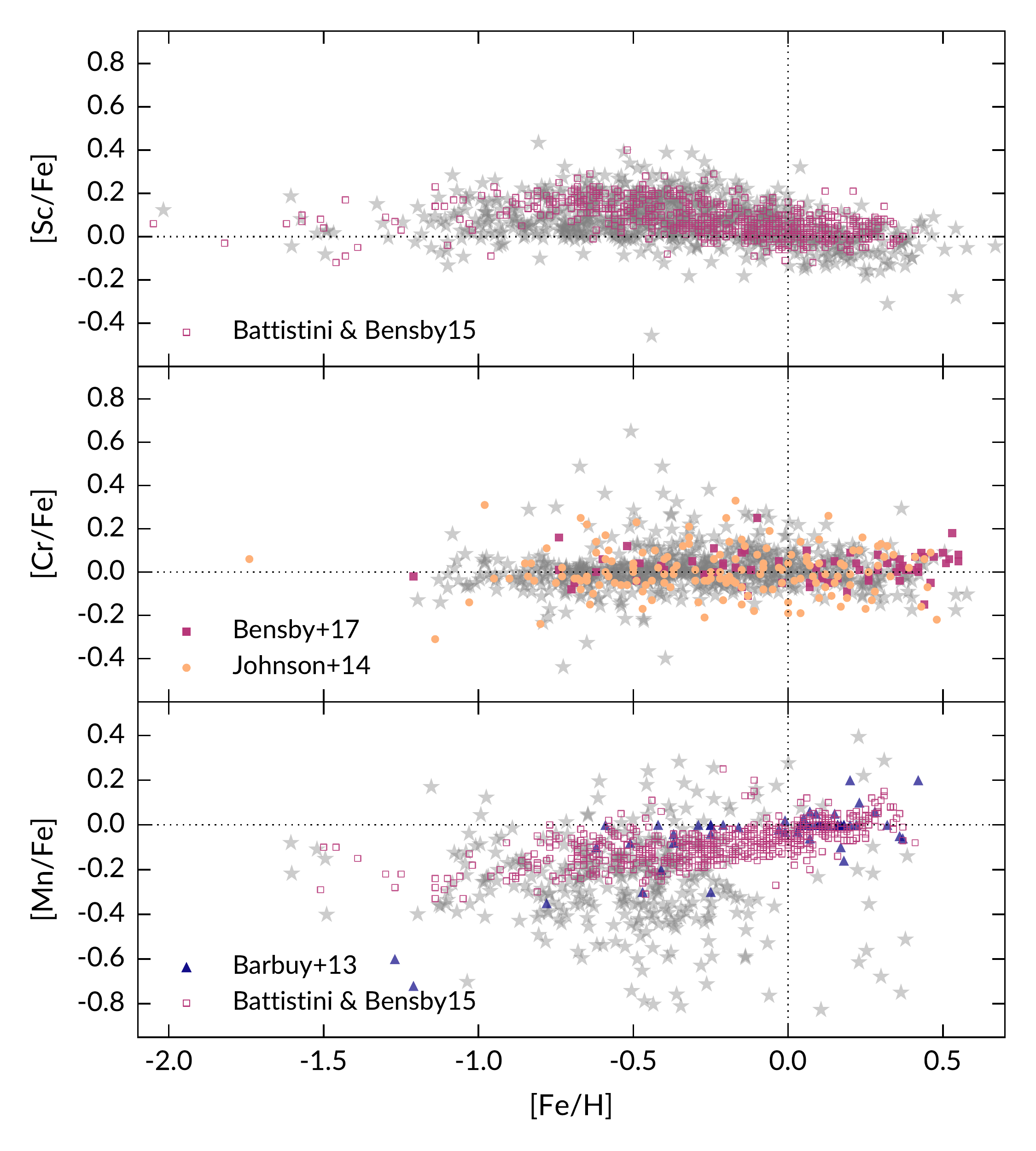}	\includegraphics[width=1\columnwidth]{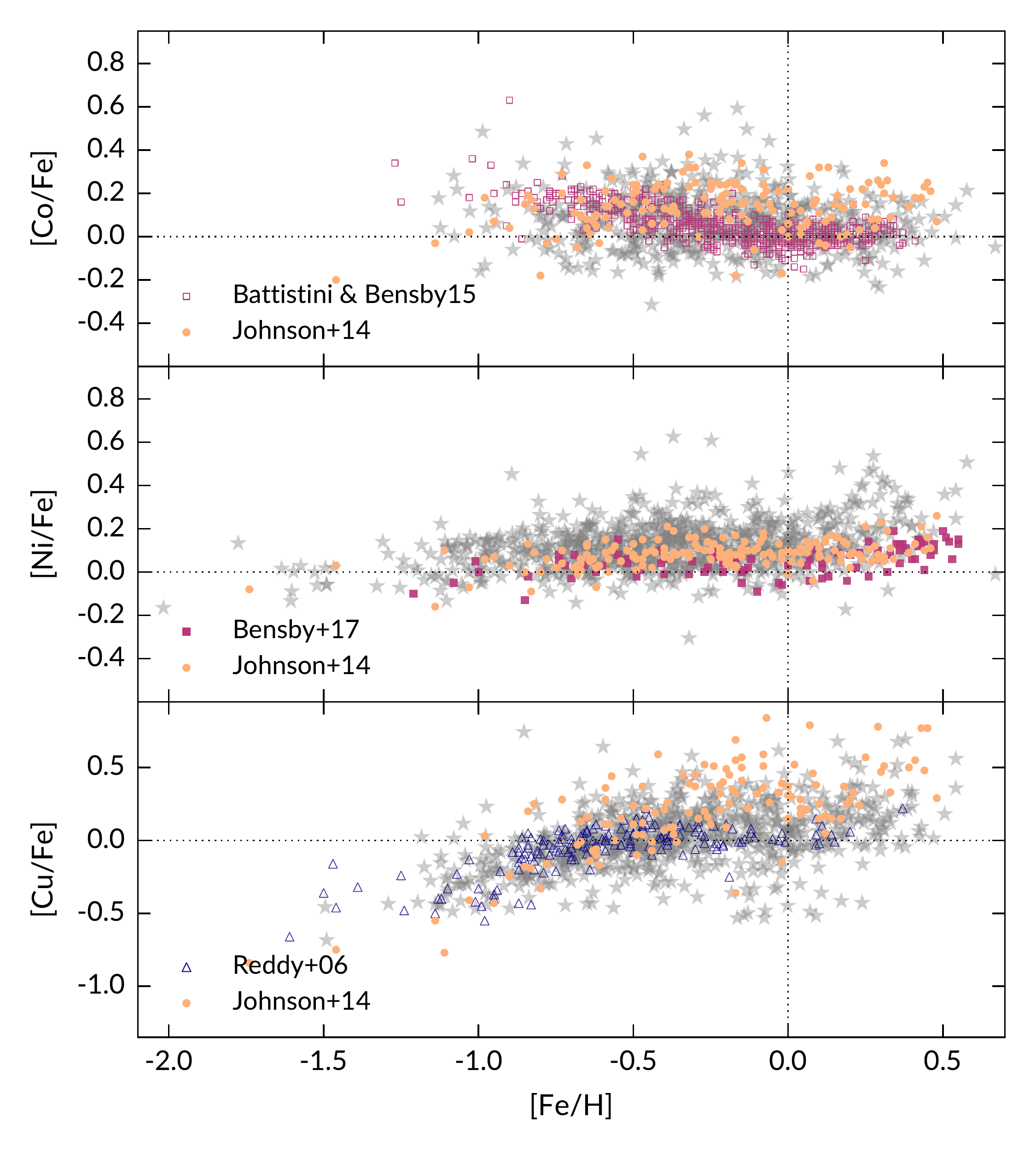}
    \caption{The [X/Fe] vs [Fe/H] trends for the iron-peak elements from this work (grey stars). Where few, or no bulge data for field stars can be compared against, data from studies of disk stars are included for reference~\protect\citep{Battistini2015,Reddy2006}.}
    \label{fig:fepeakelems}
\end{figure*}

\subsection{Neutron capture elements}

The five neutron capture abundance ratios we report here contains the $r$-process element Eu; three elements with high $s$-process contribution La, Ce, Nd and Zr~\citep{Bisterzo2014}. 

The abundance ratios of $s$-process elements La and Nd do not show strong variations with metallicity. In particular [La/Fe] remains solar at the metal-rich regime, and increases slightly for [Fe/H] $< -0.5$ (up to [La/Fe] $\approx$ 0.2). This is in line with the [La/Fe] trend observed by \cite{Simmerer2004}.~\cite{Battistini2016}, on the other hand, reported relatively under-abundant [La/Fe] for disk stars at super-solar metallicity. Our results are not in agreement with \cite{Johnson2012}, who show, on the mean, sub-solar [La/Fe] for [Fe/H] $>-0.8$. For [Nd/Fe], the trend observed in this work is similar to that of \cite{Battistini2016}: the abundance ratios increase from sub-solar at [Fe/H] $> 0$ to [Nd/Fe] $\approx 0.2$ at the metal-poor regime. \cite{Johnson2012} found the same trend, but their abundance ratios show quite large scatter around solar metallicity. The [Ce/Fe] ratios we measure remain at around solar value for [Fe/H] $< 0$. However, at super-solar metallicity, we observe enhanced cerium abundance ratios. Increasing [Ce/Fe] with metallicity could suggest a contributing blend that has not been accounted for in one for more of our Ce lines. At sub-solar metallicity, our [Ce/Fe] abundance ratios agree well with \cite{Battistini2016}. The flat [X/Fe] trends of La, Nd and Ce reflect high $s$-process contribution from low mass AGB stars~\citep{Travaglio1999,Travaglio2004,Karakas2014}. Since these stars have longer lifetimes and begin to contribute at higher metallicities, the yields of La, Nd and Ce increase with [Fe/H] and therefore we do not observe a steep decline in [X/Fe]. 

{The lanthanum, europium and zirconium abundance ratios show linear correlations with metallicity, decreasing as metallicity increases. However, the [La/Fe] slope that we observe is relatively shallow. We observe a strong slope for both europium and zirconium, but [Eu/Fe] is enhanced compared to [Zr/Fe].} This observation is broadly in agreement with both \cite{Johnson2012} and \cite{Battistini2016}. However, note that our [Zr/Fe] values show much larger scatter at super-solar metallicity, and are almost $\approx$ 0.2 dex under-abundant compared to \cite{Battistini2016}. Our [Eu/Fe] values also show much larger scatter than the literature samples. The [Eu/Fe] trend is very similar to the alpha elements and aluminium, and indicates a possible SNeII origin for this element. Our results suggest a [Eu/Fe] plateau at [Fe/H] $\lesssim$ 0.5 dex for the bulge.  It is worth noting that the decline of [Zr/Fe] with metallicity observed here (and by \citealt{Johnson2012,Battistini2016}) is steeper than the other $s$-process elements, despite Zr having similar $s$-process fraction to Nd. This may suggest a higher $r$-process contribution to zirconium yields than expected~\citep[see also][]{McWilliam2016}.

\begin{figure*}
	\centering
\includegraphics[width=1\columnwidth]{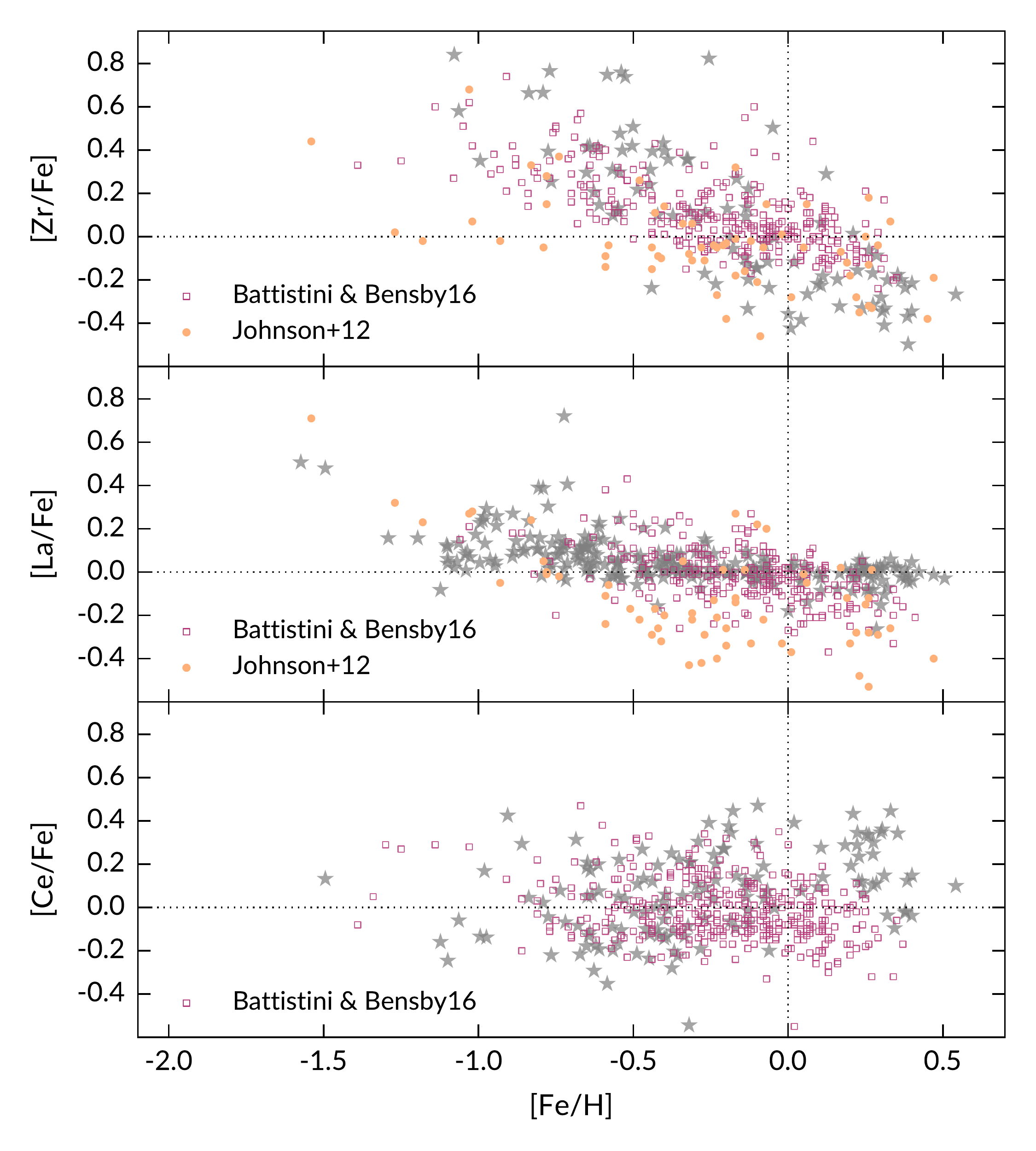}\includegraphics[width=1\columnwidth]{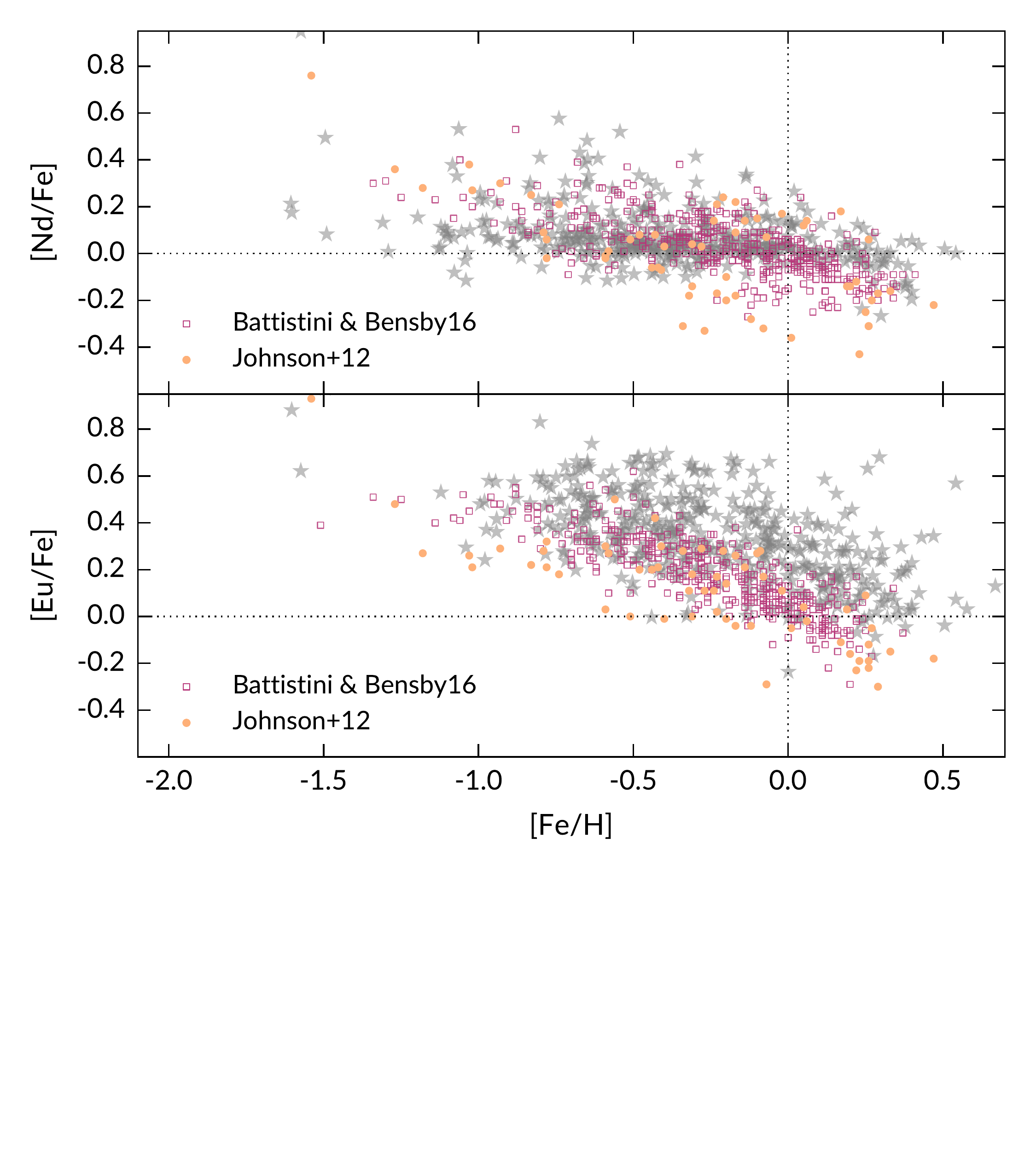}
    \caption{The [X/Fe] vs [Fe/H] trends for the neutron-capture elements from this work (grey stars). Where few, or no bulge data for field stars can be compared against, disk data from \protect\cite{Battistini2016} are included for reference.}
    \label{fig:heavyelems}
\end{figure*}

The abundance trends we observed for La, Nd and Eu agree qualitatively with~\cite{Swaelmen2016}, who reported neutron-capture abundance ratios for 56 red giants in the bulge. The [Ce/Fe] ratios from that work show very high/low values at low/high metallicities, which is not consistent with our abundance trend. Furthermore,~\cite{Swaelmen2016} reported lower [Eu/Fe] values compared to our results and the comparison samples shown in Fig \ref{fig:heavyelems}.

\section{Latitude variation and comparison to the disk and halo}  
\label{sec4}

The aim of this section is to identify and discuss the similarities and differences between the chemical evolution of the bulge and other Galactic components. Such comparison can give insights on how the bulge formed and evolved~\citep[e.g.,][]{McWilliam2016,Haywood2018}. As mentioned previously, we strive to be as consistent as possible with the GALAH survey when conducting our analysis. This is to facilitate a consistent comparison between the bulge and other Galactic components observed by GALAH. However, in this study we decided to remove certain lines from the GALAH linelist that show unexpected trends, or prominent blends. Because we used the same analysis method and atomic data, our results are still mostly consistent with GALAH. For some elements, however, the removal of these lines caused discrepancies between this work and the GALAH survey. For example, we observe systematic differences in [Sc/Fe] abundance ratios, where our results are $\approx$0.2 dex lower than GALAH values, as shown in Fig~\ref{fig:galah-comp}. The likely reason for this is our removal of a scandium line that showed very enhanced abundance ratios at all metallicities, and a strong increase in [Sc/Fe] with with increasing [Fe/H]. We note that the majority of GALAH dwarf stars do not deviate significantly from the trend reported by \cite{Battistini2015}, suggesting that the issue mainly manifests in giants (see Fig. 23 of \citealt{Buder2018}). We therefore cannot provide conclusions for [Sc/Fe] in terms of its connection to other Galactic components. Furthermore, we do not provide a comparison for the element Nd, as these abundance ratios were not released as part of GALAH Data Release 2.

In the sections below, we have selected GALAH disk and halo giants with similar stellar parameters to our bulge sample for comparison. We also used the same definition as in Paper I to assign disk/halo membership. The median bulge abundance trends and standard errors were calculated for [Fe/H] bins that are $\approx$0.2 dex wide. Here, data points that lie more than two sigma from the median were omitted so that metallicity bins with few data points are not affected by outliers. The median bulge abundance trends are also used to assess whether or not different bulge populations show latitude-dependent variations. Mn, Zr and Ce are excluded entirely from this exercise because we have too few data points to compute mean abundance trends. We note that Zr and Ce abundance ratios were also not available in the GALAH dataset.  

\subsection{Light elements} 

For both Na and Al, we do not observe significant systematic latitude variations at fixed metallicity given the standard errors. The two disk populations (low and high-$\alpha$) are not distinguishable by their [Na/Fe] abundances, but some distinction can be seen in [Al/Fe]. The high-$\alpha$ population is also enhanced in [Al/Fe] compared to the low-$\alpha$ population. {\cite{Fulbright2007} also measured similar [Al/Fe] ratios to ours for bulge stars at [Fe/H] $\leq -0.2$.} 

{Our comparison shows that the disk and bulge basically have the same [Na/Fe] trends for [Fe/H] $\geq -1$ \citep[see also][]{Alves-Brito2010}}. At metallicities lower than $-1$, there is a difference between the bulge and high-$\alpha$ disk/halo: the bulge [Na/Fe] remain above solar, meanwhile [Na/Fe] decrease to sub-solar values for the high-$\alpha$ disk and halo. In addition, we find that the aluminium abundances in the bulge is generally enhanced compared to the low-$\alpha$ disk, and is consistent with the behaviour of the high-$\alpha$ disk. From [Fe/H] $\approx -0.8$, the bulge [Al/Fe] trends lie above the high-$\alpha$ disk. The few halo stars in the GALAH sample show very large scatter in [Al/Fe]. {We note that there are some caveats concerning the light element comparison between the HERBS and GALAH samples (refer to Appendix \ref{herbs-galah-comp}), but they do not change our conclusions here.}

It has been confirmed by many authors that aluminium is produced in massive stars, and the [Al/Fe] trend closely resemble that the alpha elements. The so-called `zig-zag' [Na/Fe] trend can be understood if Na is produced in massive stars, but the yield is metal-dependent (i.e., increasing with [Fe/H])~\citep{McWilliam2016,Bensby2017}. If both aluminium and sodium are primarily produced in SNeII, then it appears that the metal-poor bulge population ([Fe/H] $< -0.8$) contains excess SNeII ejecta compared to the disk and halo. 

{The [Na/Fe] and [Al/Fe] ratios we measured do not exhibit anti-correlation behaviour with [O/Fe] and [Mg/Fe], as often observed in globular clusters. We could expect to see stars with globular cluster chemistry in our sample, since \cite{Schiavon2017,Fernandez2017} found nitrogen-rich bulge stars that could have originated from a dissolved globular cluster in APOGEE data. It is unclear, however, if these stars could be distinguished in our abundance space: While the [Al/Fe]-[N/Fe] correlation was observed, the [Al/Fe]-[Na/Fe] correlation was not confirmed by \cite{Schiavon2017}.} 

\subsubsection{\emph{[Na/Fe]}, \emph{He}-enhancement and the double red clump}

The double red clump (double RC), or the presence of a faint and bright red clump population was confirmed by multiple authors from photometric surveys~\citep[e.g.,][]{McWilliam2010, Nataf2015}. The double RC is typically associated with the X-shaped morphology of the bulge, but this is not without contention. \cite{Lee2015} and \cite{Joo2017} argue that double clump can arise from two populations with different helium content, such that the helium-enhanced population manifests as a brighter red clump branch. This would be similar to multiple stellar populations with distinct helium abundances observed in globular clusters (GCs)~\citep[e.g.,][]{D'Antona2010,Dupree2011}. Because helium-enhanced populations in GCs also show strong sodium enhancement~\citep[e.g.,][]{Carretta2010}, one would expect to observe the same [Na/Fe] enhancement in the bulge. At high latitudes a long the minor axis (e.g. $b = -10^\circ$), we should also see a higher fraction of sodium-rich stars than at lower latitudes according to the model of ~\cite{Lee2015}. In contrast, our results here show that [Na/Fe] is very much uniform at all latitudes. \cite{Joo2017} also cited the larger spread in bulge [Na/Fe] measurements~\citep{Johnson2014} compared to the disk/halo as evidence that the bulge contains more helium-rich stars. But this conclusion is based on multiple different studies, which could have large systematic offsets and differences in their intrinsic precision. Both \cite{Bensby2017} and this work have shown that the spread in bulge and disk sodium abundances are comparable for stars that are homogeneously analysed. In theory, helium enhancement similar to the phenomenon observed in globular clusters may be able to explain the split red clump in the bulge, but this explanation is not supported by our [Na/Fe] measurements. {However, it is still possible for the bulge RC to contain populations with different He content.}  

\subsection{Iron-peak elements} 
\begin{figure*}
	\centering
    \includegraphics[width=1\columnwidth,trim={0.8cm 0.85cm 0.3cm 0.85cm},clip]{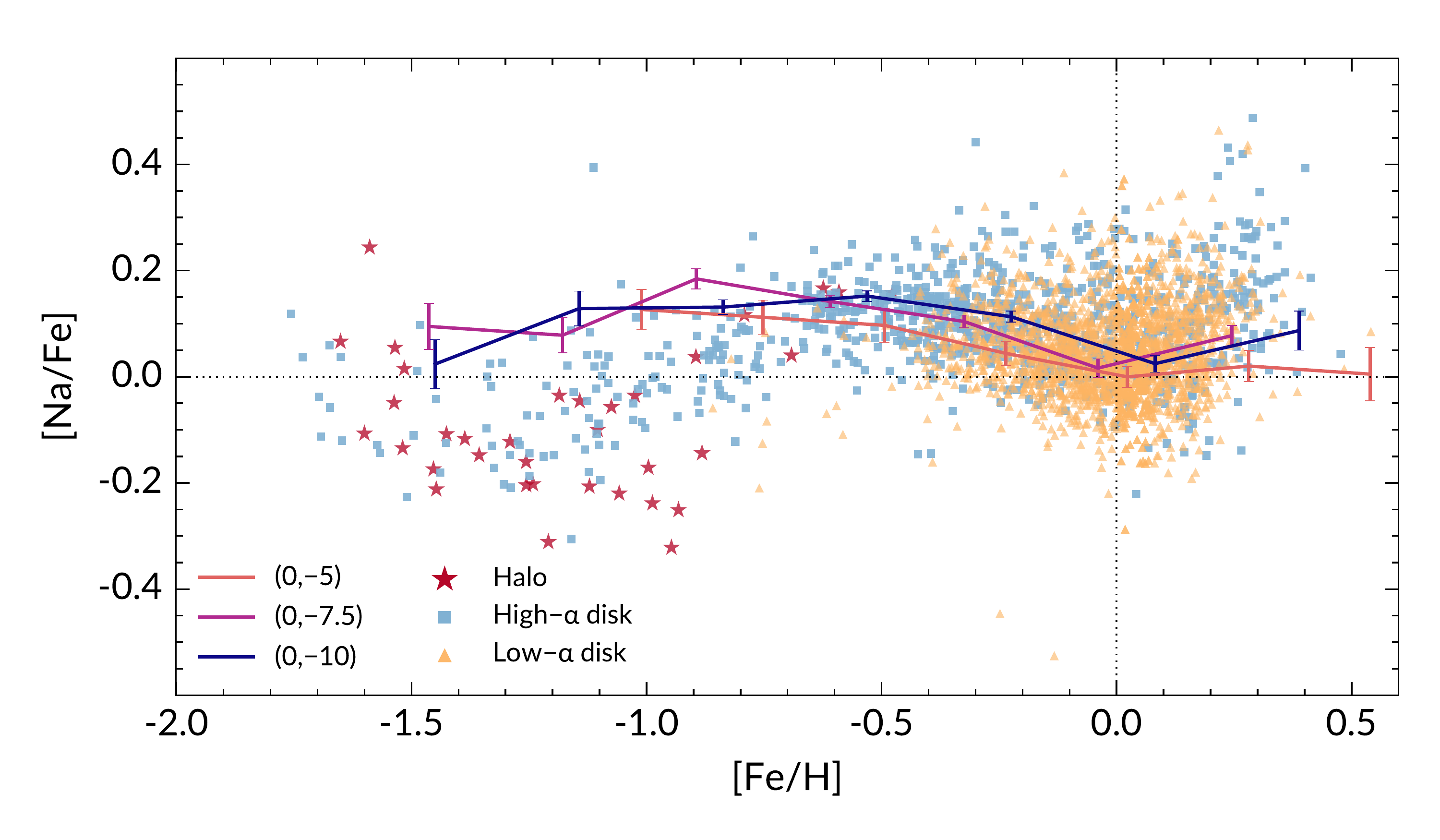}
	\includegraphics[width=1\columnwidth,trim={0.8cm 0.85cm 0.3cm 0.85cm},clip]{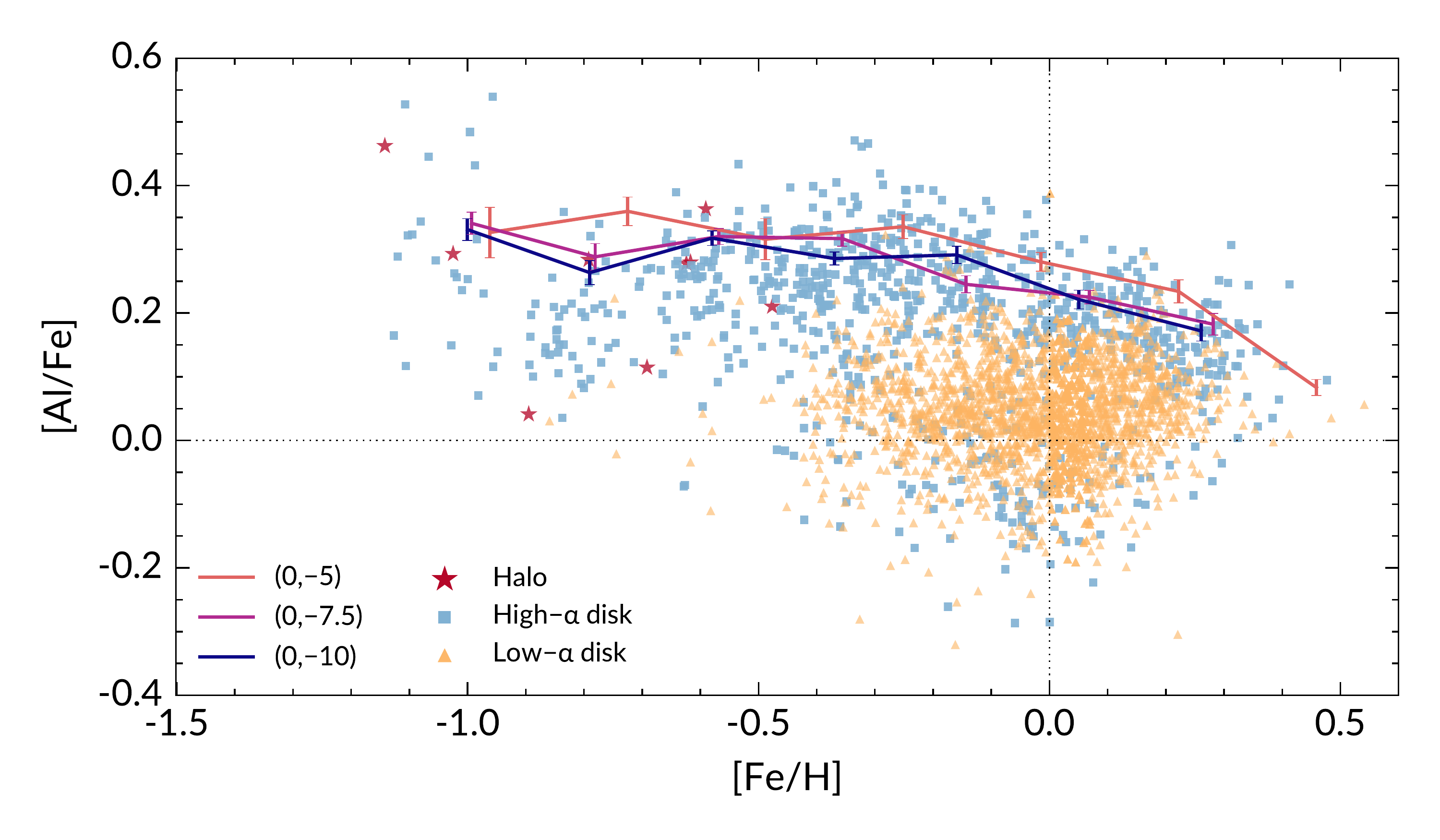}
    \includegraphics[width=1\columnwidth,trim={0.8cm 0.85cm 0.3cm 0.85cm},clip]{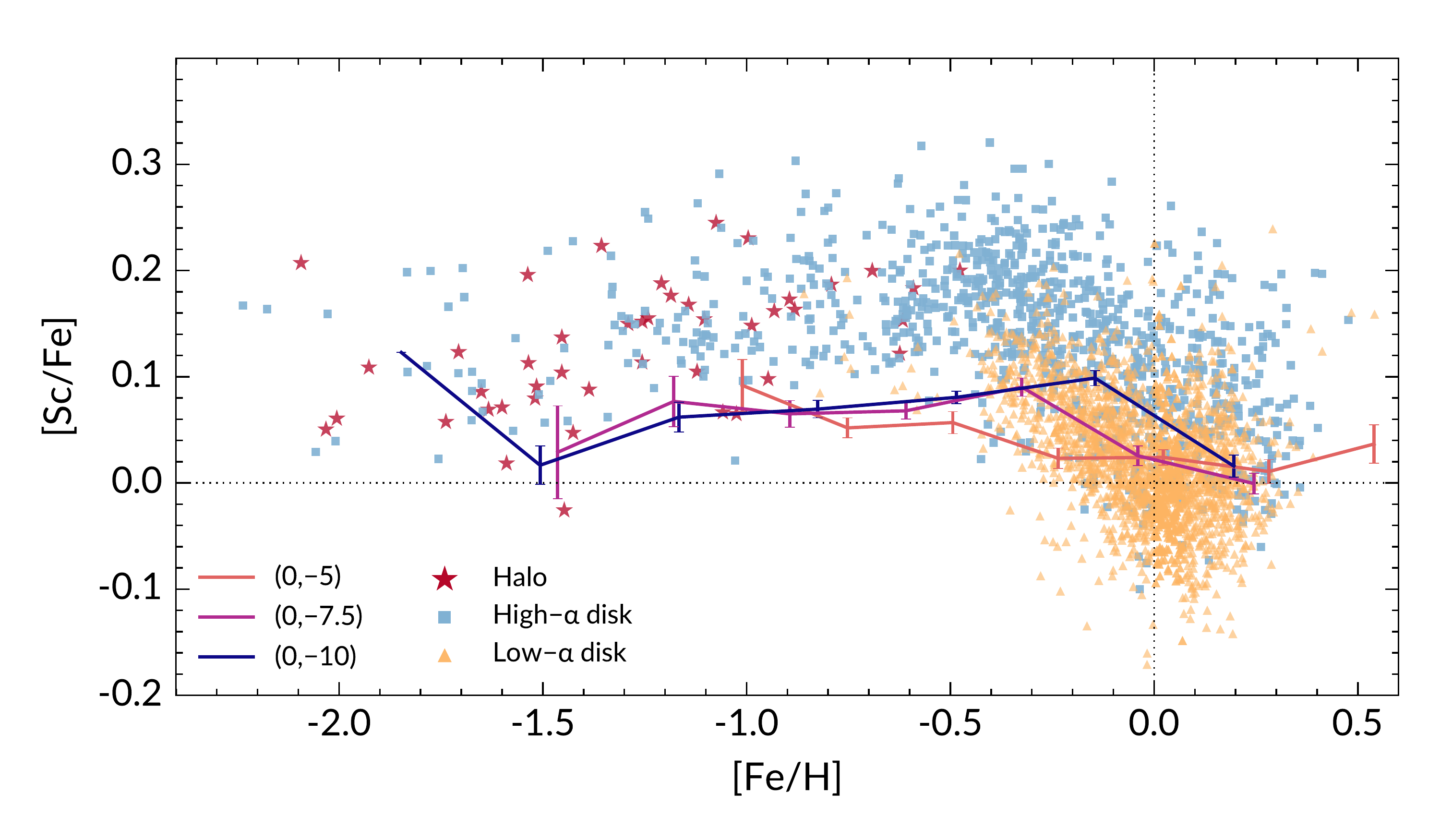}
    \includegraphics[width=1\columnwidth,trim={0.8cm 0.85cm 0.3cm 0.85cm},clip]{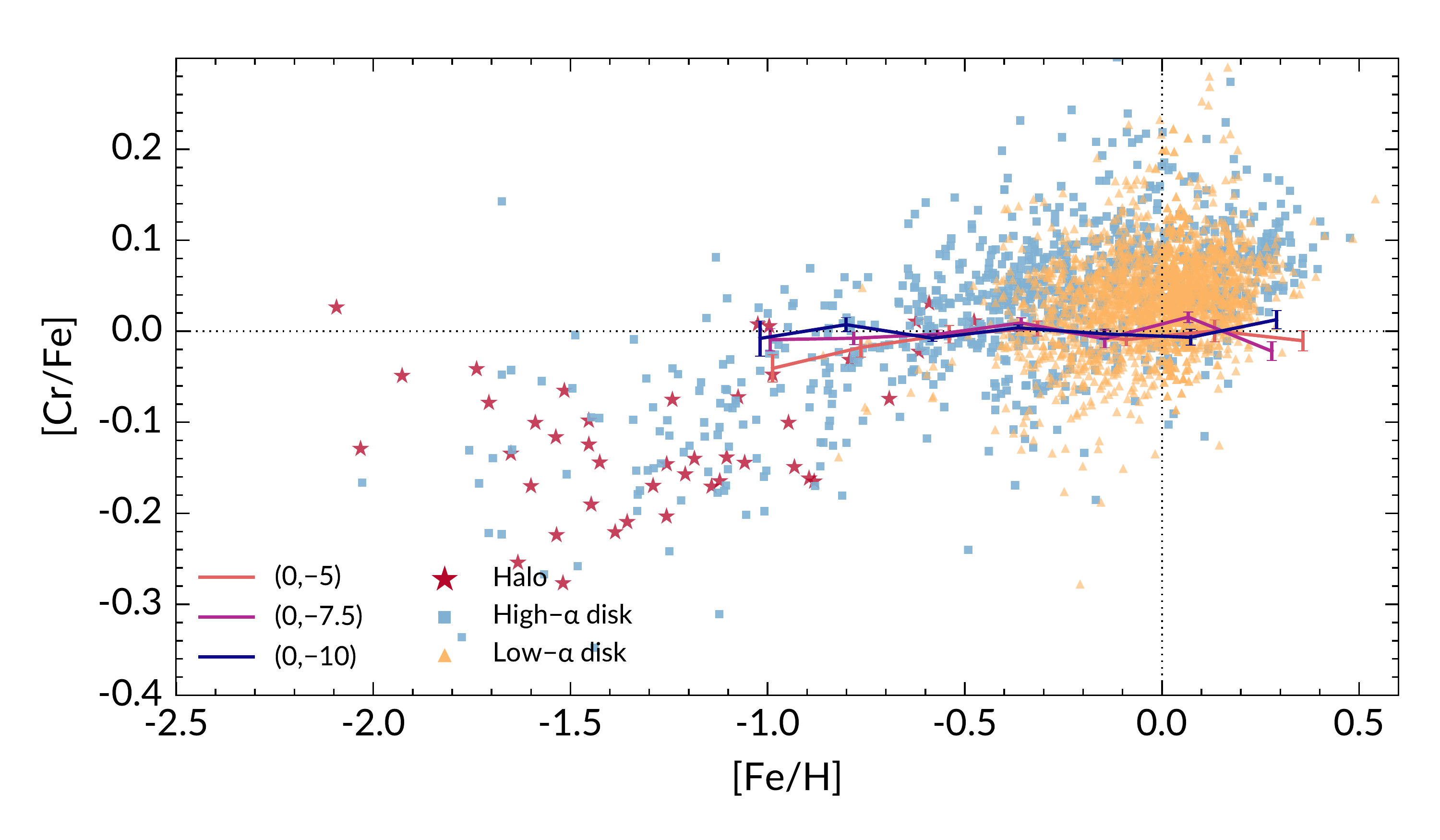}
    \includegraphics[width=1\columnwidth,trim={0.8cm 0.85cm 0.3cm 0.85cm},clip]{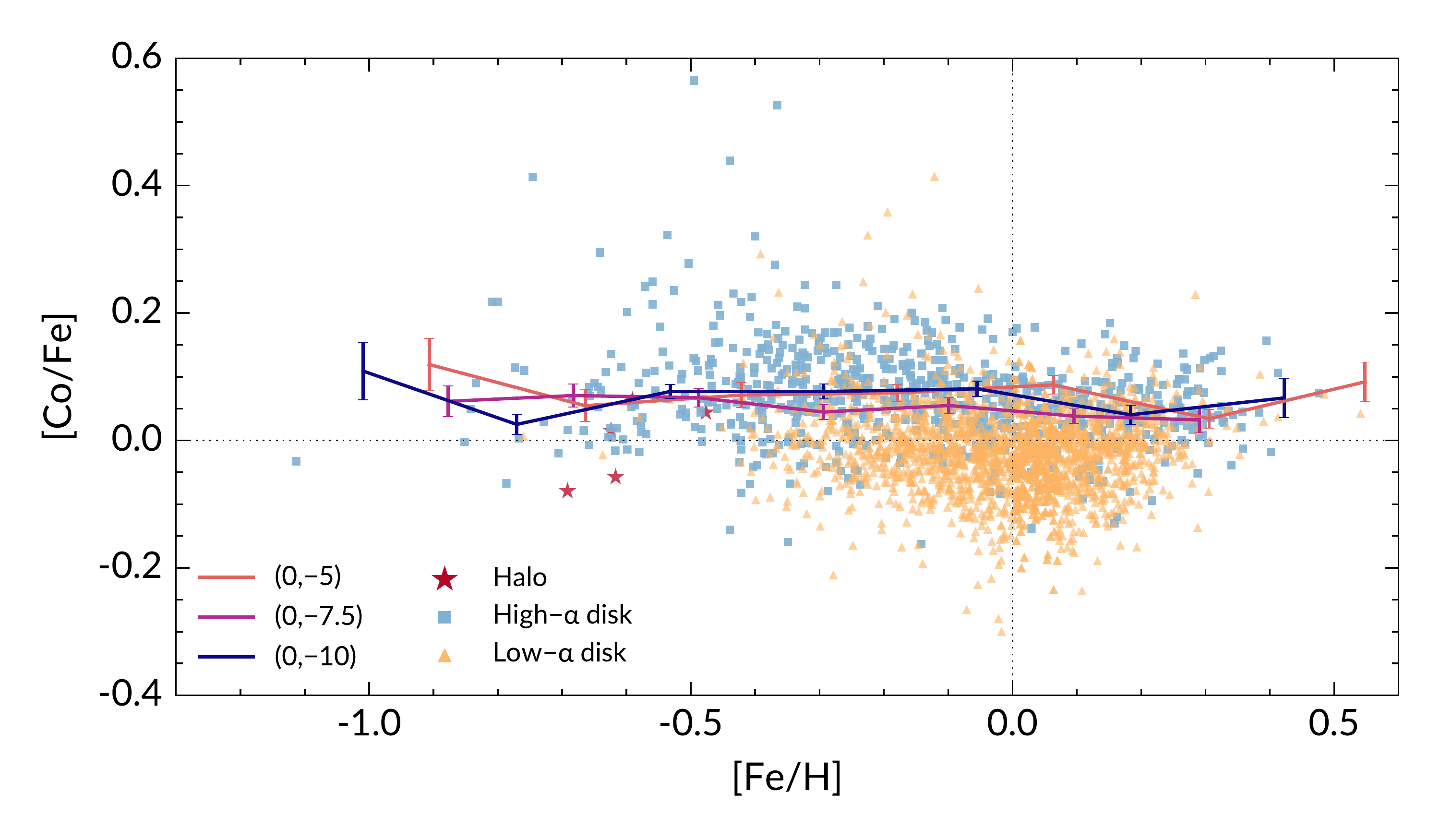}
	\includegraphics[width=1\columnwidth,trim={0.8cm 0.85cm 0.3cm 0.85cm},clip]{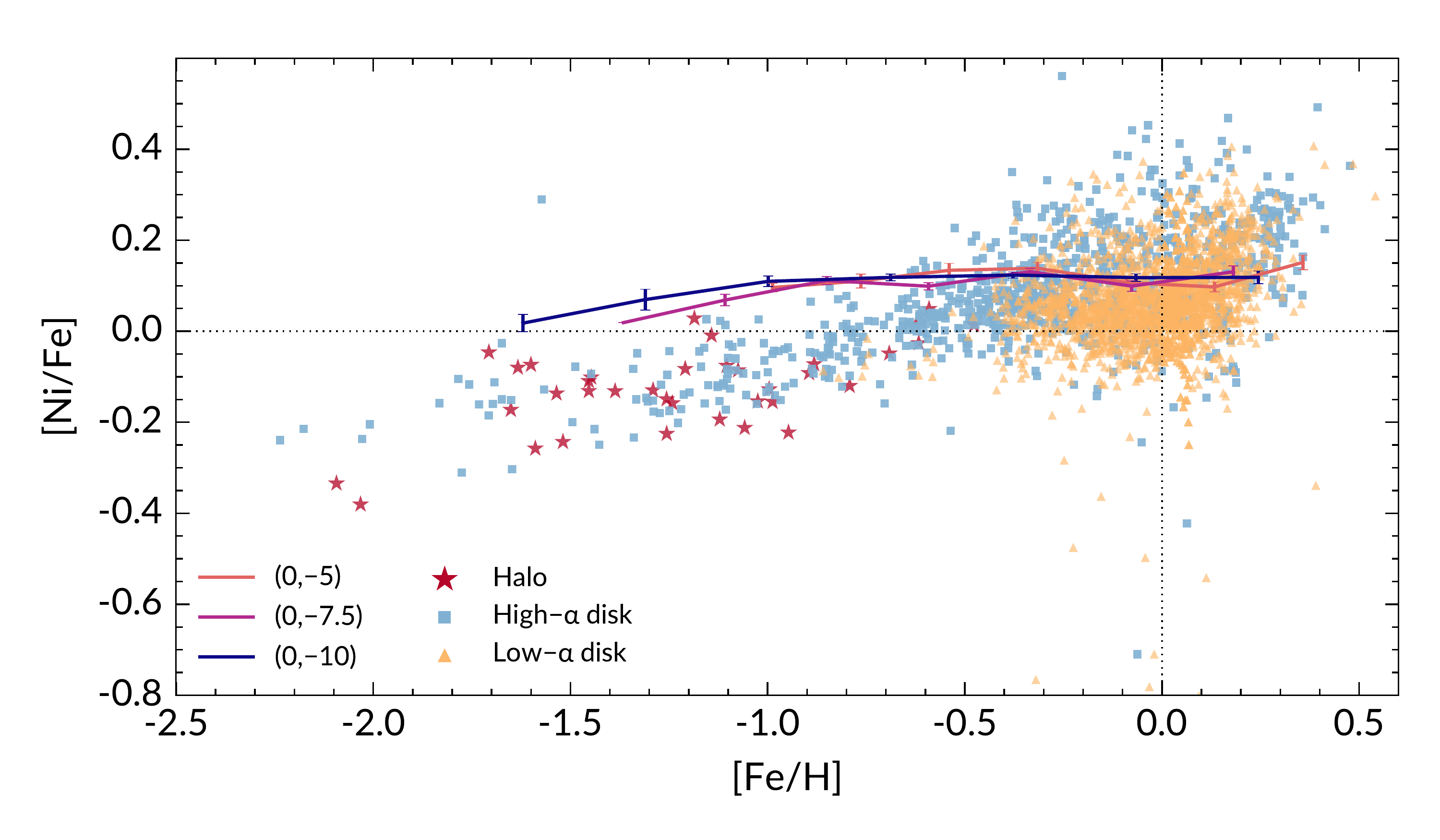}
	\includegraphics[width=1\columnwidth,trim={0.8cm 0.85cm 0.3cm 0.0cm},clip]{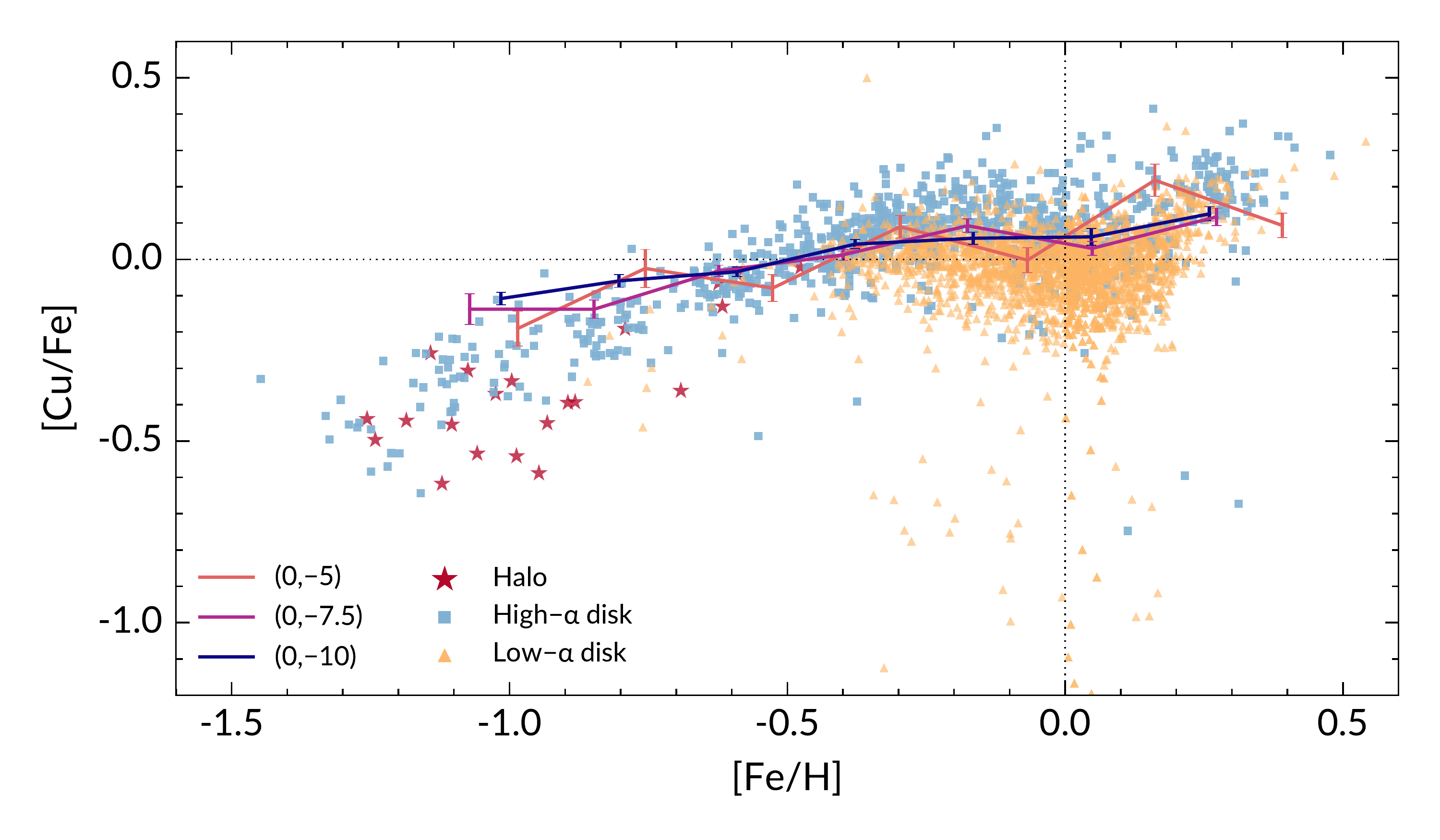}
    \includegraphics[width=1.02\columnwidth,trim={0.8cm 0.85cm 0.3cm 1cm},clip]{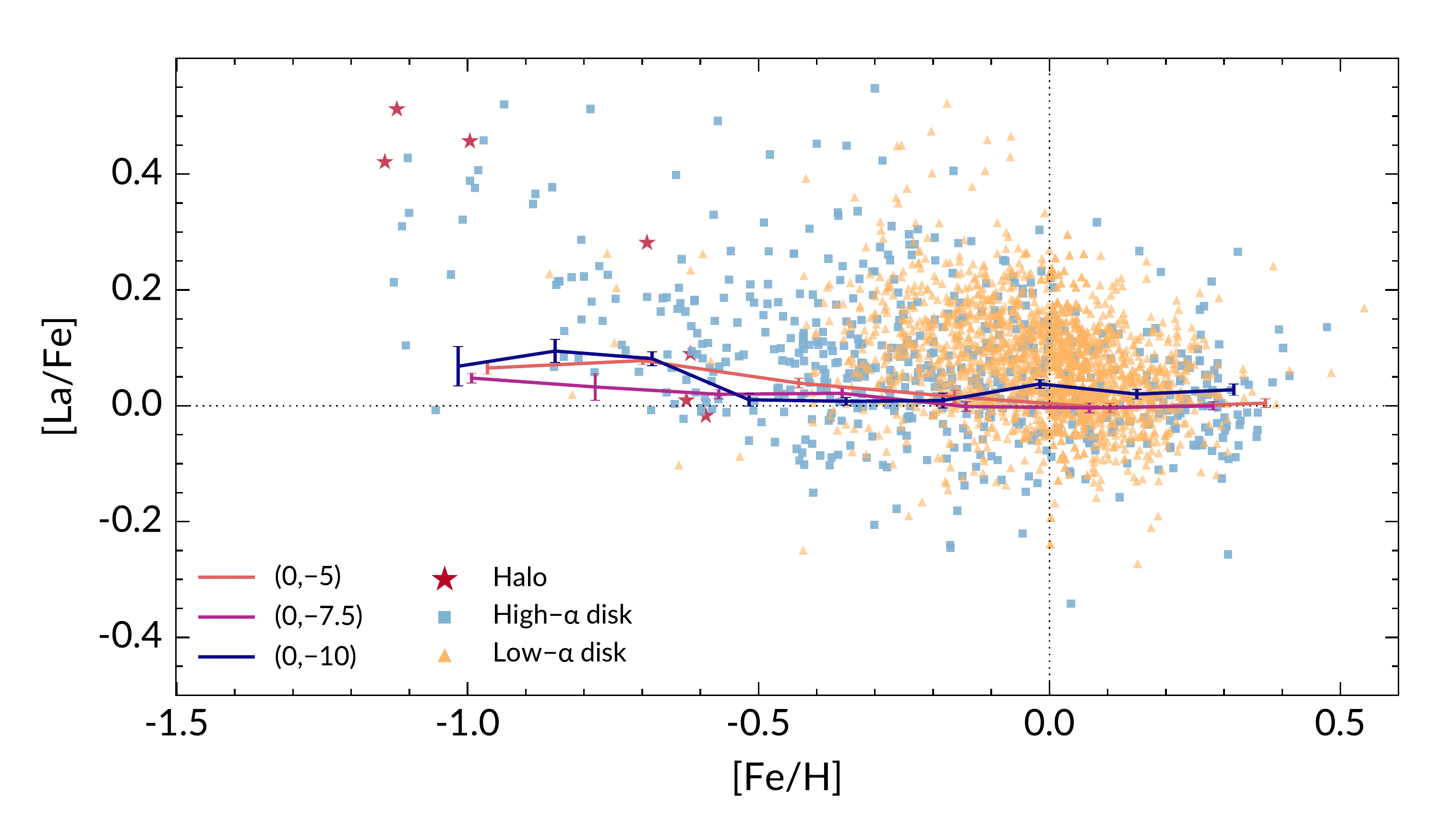}
    \includegraphics[width=1\columnwidth,trim={0.8cm 0.85cm 0.3cm 0.85cm},clip]{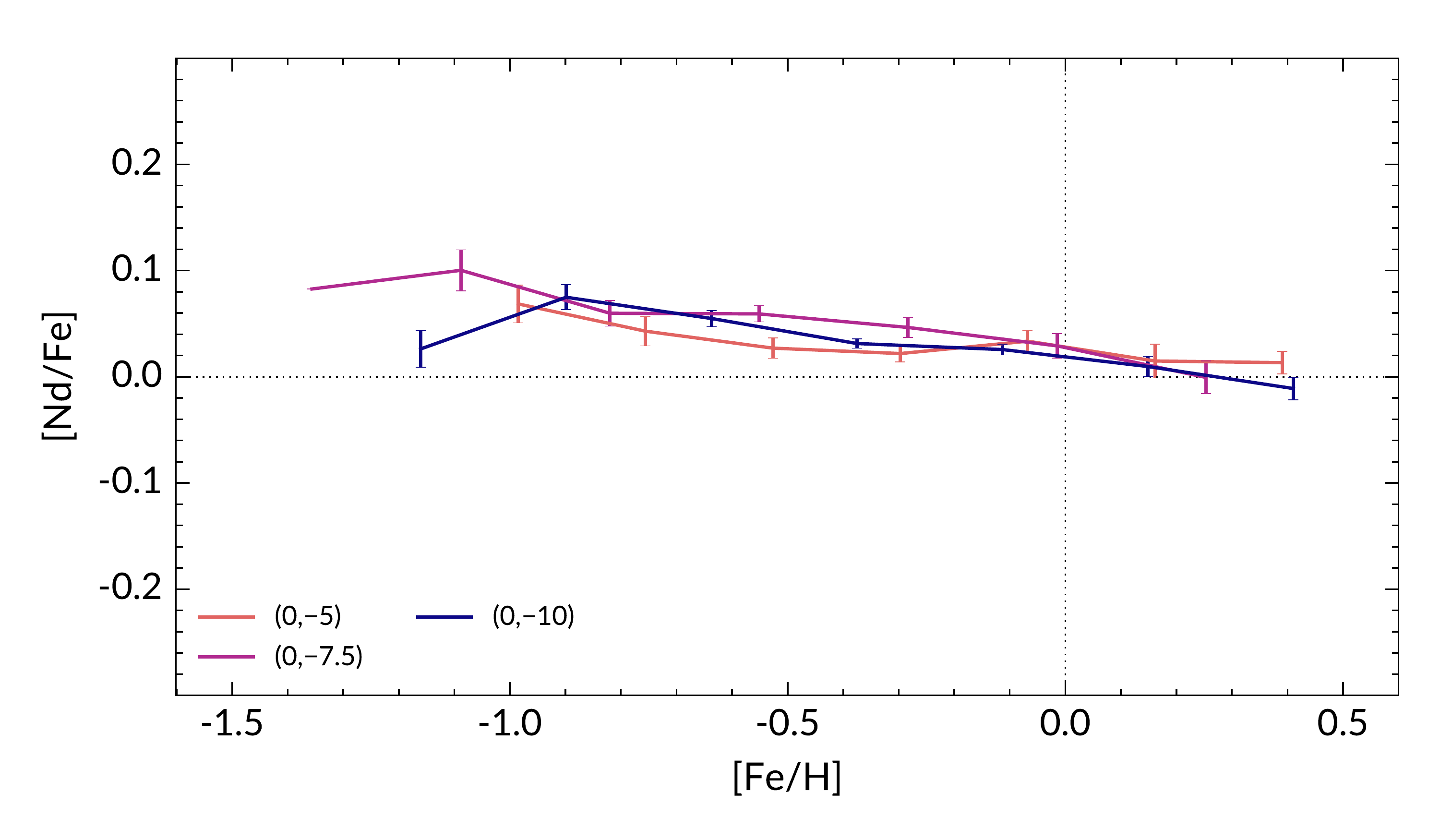}
	\includegraphics[width=1\columnwidth,trim={0.8cm 0.85cm 0.3cm 0.85cm},clip]{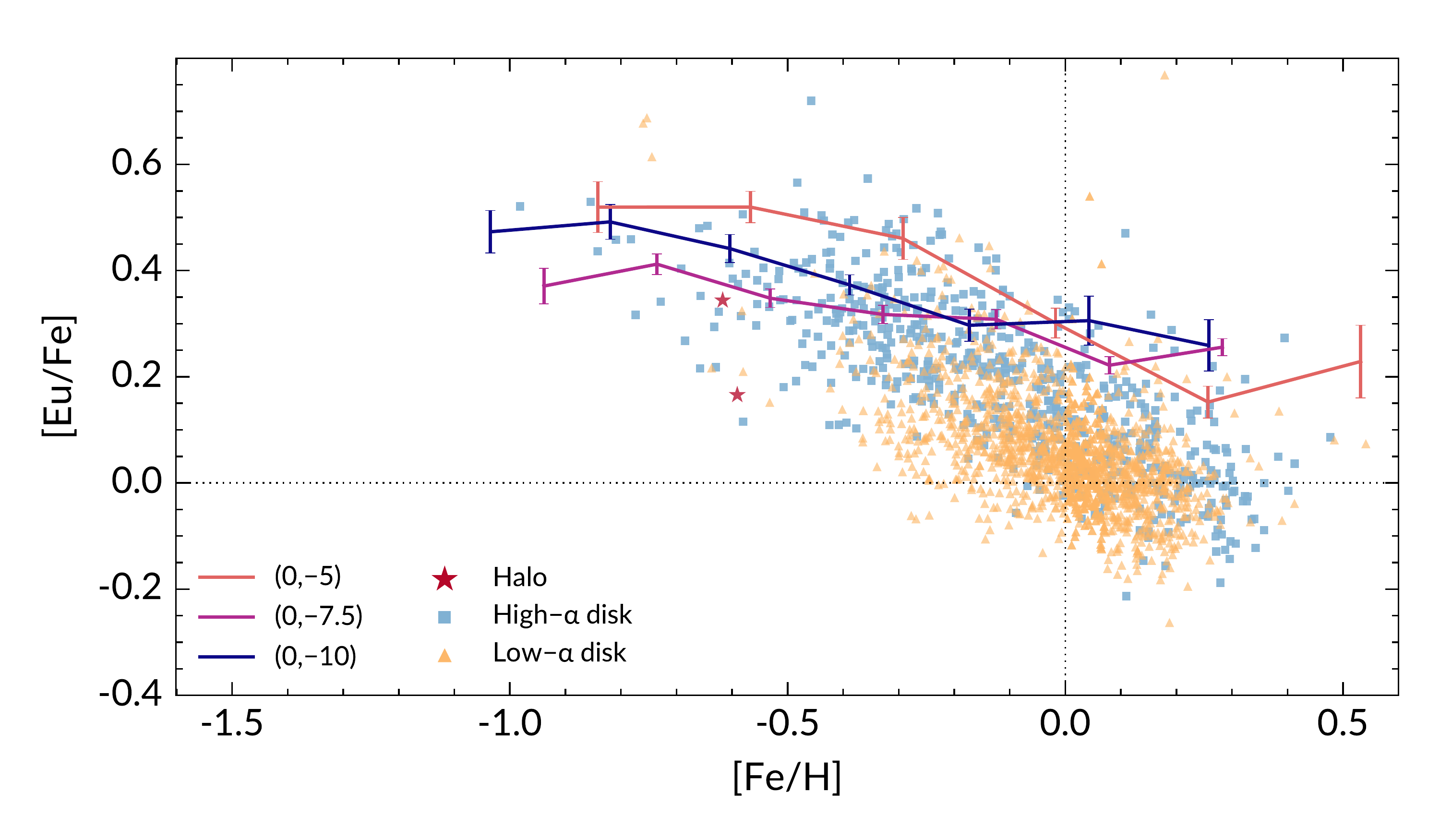}
	\caption{Comparison of the bulge and GALAH disk/halo sample for all available elements. The solid lines indicate median bulge trends at three minor-axis latitudes. While the GALAH [Sc/Fe] ratios are plotted here, we \emph{do not} compare our results with GALAH due to systematic offsets (see text for details).}
	\label{fig:galah-comp}
\end{figure*}
Within errors, the iron-peak elements mostly show uniform abundance ratios as a function of latitude, as shown in Fig~\ref{fig:galah-comp}. Scandium exhibits slight latitude variations around solar metallicity, where field $(0, -10)$ has higher [Sc/Fe] than the other fields. {However, this may be due to the small number of stars in each bin at high metallicities}. We do not observe differences between bulge and disk abundances for cobalt\footnote{The high-$\alpha$ and low-$\alpha$ disks differ in their mean [Co/Fe] by 0.05 dex, smaller than the typical abundance uncertainties (0.08 dex). Because the two disk populations overlap substantially, and extend the same range, we did not discuss them separately.}. Both the bulge and disk show a rather flat [Co/Fe] trend, with a modest increase at low metallicities. Although we show the GALAH [Sc/Fe] ratios against ours, the large systematic difference prevents us from making a meaningful comparison. 

The GALAH survey was able to measure [Cr/Fe] to much lower metallicity than we could (due to better SNR of GALAH stars), but with larger scatter. For all stars with $< -1$ [Fe/H] $< 0.2$, where we could measure [Cr/Fe], the bulge abundance ratios follow the same trend as the disk, which remains constant at all metallicities. However, GALAH disk stars may have slightly enhanced mean [Cr/Fe] (by $\approx$0.05 dex) compared to bulge stars. This perhaps indicate a small zero point error in GALAH results. We note that the GALAH [Cr/Fe] ratios begin to decrease with metallicity for [Fe/H] $< -1$, and at [Fe/H] $\approx -1$, the mean GALAH [Cr/Fe] seems to be slightly lower than ours. However, this conclusion is less certain due to the lack of bulge data points at this metallicity regime. 

For nickel, the bulge abundance ratios are in agreement with that of the disk. There is a difference compared to GALAH disk/halo stars in the metal-poor regime ([Fe/H] $< -0.8$). Here the bulge is enhanced in [Ni/Fe] by up to 0.2 dex compared to the disk and halo. This enhancement at low metallicity is qualitatively similar to the behaviour of [Na/Fe] we observed in the previous section. As \cite{McWilliam2016} noted, enhanced [Ni/Fe] in the metal-poor bulge population could indicate more SNeII material in their birth environment. Some halo stellar populations that are deficient in SNeII products, such as the alpha elements, also show relatively lower [Ni/Fe] ratios~\citep[e.g.,][]{Nissen2010}. {From $-0.8 \lesssim$ [Fe/H] $\lesssim -0.5$, the bulge has slightly enhanced [Ni/Fe] compared to the disk (by 0.05 dex). But this is similar to the typical abundance ratio uncertainties of $\approx$0.04 dex.}

The bulge copper abundance ratios follow the same trend as the GALAH sample. For all stars with [Fe/H] $\geq -0.8$, there is no noticeable difference between the mean [Cu/Fe] of the bulge and the disk. In addition, both populations clearly show increasing [Cu/Fe] at super-solar metallicity. At $-1 \lesssim $ [Fe/H] $\lesssim -0.8$, it appears that the bulge [Cu/Fe] median trends lie just above disk/halo stars. {However, our abundance pipeline underestimates [Cu/Fe] by 0.12 dex compared to GALAH (see Appendix \ref{herbs-galah-comp}). Taking this offset into account, the bulge likely has enhanced [Cu/Fe] compared to the disk, especially at [Fe/H] $\lesssim -0.8$.} The variation of [Cu/Fe] with metallicity suggests that copper is produced via the weak $s$-process in massive stars, and its yield increases with metallicity~\citep[e.g.,][]{Romano2007}. As mentioned previously, \cite{Johnson2014} reported enhanced [Cu/Fe] around [Fe/H] $\geq -0.5$. This lead \cite{McWilliam2016} to tentatively conclude that the bulge may have higher copper yields and therefore higher star formation rate than the local disk. {Although we do not weaker [Cu/Fe] enhancement in the bulge compared to \cite{Johnson2014}, our observation corroborates this finding.}

\subsection{Neutron capture elements}

\begin{figure}
	\centering
	\includegraphics[width=1\columnwidth,trim={0.8cm 0.8cm 0cm 0.8cm},clip]{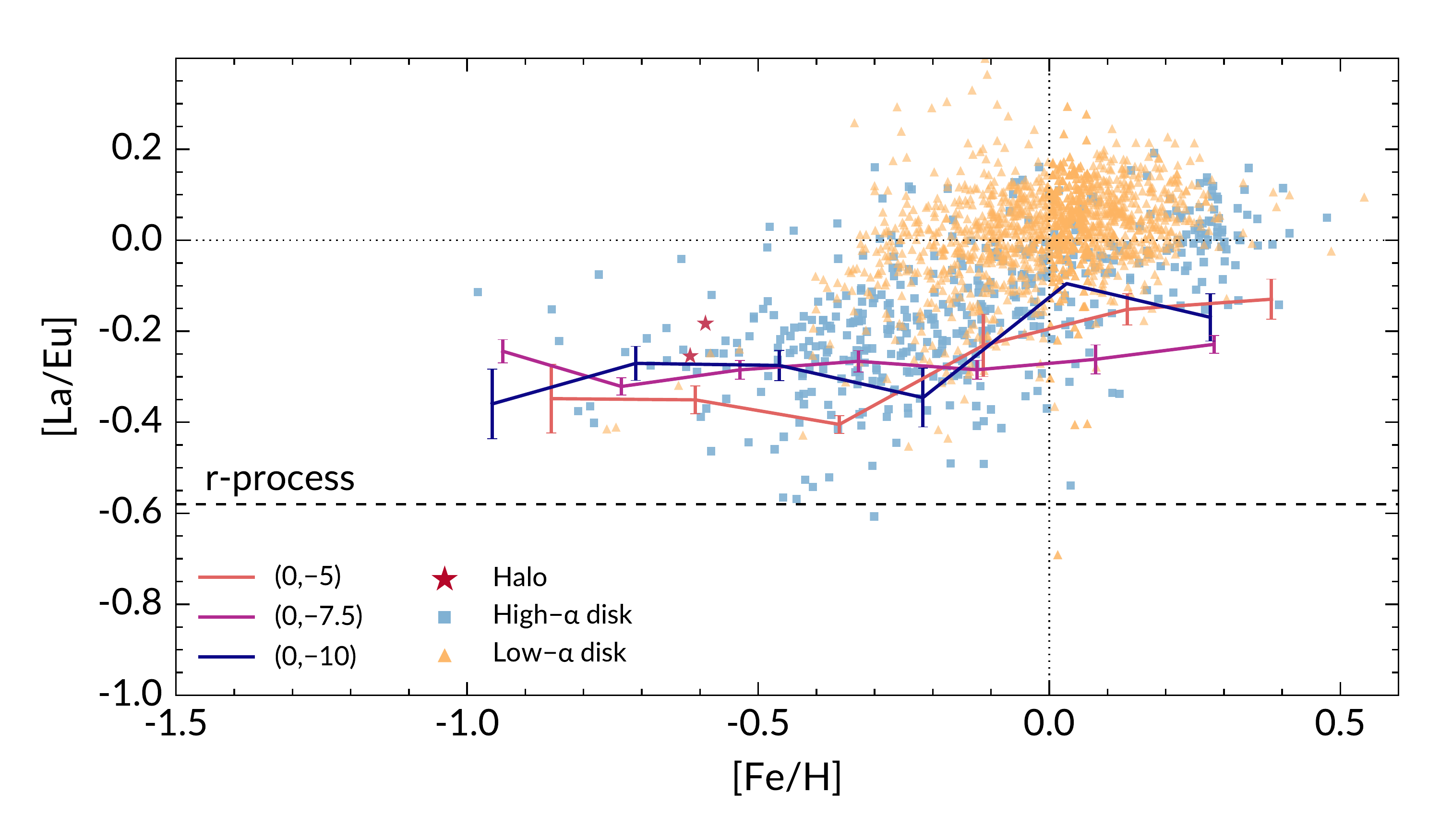}
    \caption{The comparison of $s$/$r$-process ratio in the bulge and disk/halo. The solid lines indicate median bulge trends at three minor-axis latitudes. The dashed line shows the pure $r$-process ratio from~\protect\cite{Bisterzo2014}.}
    \label{fig:heavyratio}
\end{figure}

As can be seen in Fig~\ref{fig:galah-comp}, no consistent trend with latitude is observed for La, Nd and Eu. We could not compare our [Nd/Fe] values with GALAH because they were not available as part of the survey data release. However, our [La/Fe] trend seems to be consistent with the GALAH disk sample. Both the disk and bulge show flat [La/Fe] with metallicity, but the disk has larger scatter. We also note that at sub-solar metallicity, the disk [La/Fe] is $\approx 0.1$ dex higher than the bulge. In  this case we used the same La lines as GALAH, so there are no systematic offsets (Appendix \ref{herbs-galah-comp}). The higher [La/Fe] abundance ratios in the disk could indicate more efficient $s$-process compared to the bulge. There are too few halo data points for us to make a comparison, and the scatter in [La/Fe] is very large for halo stars.

For the $r$-process element Eu, our comparison with the GALAH sample indicates that the bulge is enhanced in [Eu/Fe] compared to the disk for [Fe/H] $\gtrsim -0.4$. At lower metallicities, the disk sample becomes quite sparse and the bulge [Eu/Fe] begins to show very large scatter. This is probably the reason why for all stars with [Fe/H] $\lesssim -0.4$, we can find no difference between the disk and bulge samples. We only used one of the two Eu lines in the GALAH linelist, but the line that we discarded is very weak, and if anything this line would have contributed even higher Eu abundances. 

In Fig~\ref{fig:heavyratio}, we show the $s$/$r$-process ratio for our bulge sample and the GALAH disk/halo sample using [La/Eu] abundance ratios. Overall we observe a consistent trend with previous works, that {the [La/Eu] ratio in both the disk and bulge increase as a function of metallicity~\citep{McWilliam2010,Johnson2012,Swaelmen2016}}. However, the bulge [La/Eu] ratios are markedly lower than in the disk (see Table \ref{table:rdiff}), and are closer to the pure $r$-process ratio from~\cite{Bisterzo2014}. The bulge is more similar to the high-$\alpha$ disk population at sub-solar metallicity. From solar metallicity and above, the GALAH sample consists of mostly low-$\alpha$ disk stars, which appears to have higher $s$-process contribution. However, we should mention that \cite{Battistini2016} found no distinction between their high and low-$\alpha$ disk population, and their [La/Eu] ratios are similar to that of our bulge sample. 

\begin{table}
	\caption{The median of disk/halo and bulge median for [La/Eu] abundance ratios at different metallicity bins. See text for details.}
	\label{table:rdiff}
	\begin{tabular}{lll}
		\hline 
        Metallicity range & Bulge median & Disk median\\
        \hline
         \ \ 0.2 $<$ {[Fe/H]} $<$ 0.4 & $-0.16 \pm 0.02$ &  \ \ $0.02 \pm 0.01$\\
		 \ \ 0.0 $<$ {[Fe/H]} $<$ 0.2 & $-0.26 \pm 0.04$ &  \ \ $0.034 \pm 0.003$\\
        $-0.2 <$ {[Fe/H]} $<$ 0.0 & $-0.30 \pm 0.03$ & $-0.017 \pm 0.005$\\
        $-0.4 <$ {[Fe/H]} $< -0.2$ & $-0.28 \pm 0.03$ & $-0.12 \pm 0.01$\\
        \hline
	\end{tabular}
\end{table}
\section{Conclusion}

As part of the HERBS survey, this work reports abundances for 13 elements in the light, iron-peak and neutron-capture families, in addition to the five alpha elements reported in Paper I. For most of the elements, we can measure abundances to $0.5 \lesssim$ [Fe/H] $\lesssim -1.5$. We assessed the abundance trends at different latitudes along the minor axis and compared the chemistry of bulge stars with disk/halo stars from the GALAH survey. In addition to comparing only stars with similar parameters, the same atomic data and analysis method is used in this work and GALAH. This removes the major systematic offsets between the two sets of results.

In general, we found that the [X/Fe] vs [Fe/H] trends follow that of previous studies (of both disk and bulge stars) at similar or higher resolving power. {For the bulge in particular, we did not observe enhanced [Co/Fe] compared to the disk, nor sub-solar [La/Fe] and [Zr/Fe] at [Fe/H] $> -0.5$ as did \cite{Johnson2014}}. {The majority of our iron-peak elements agree quite well with abundances from ~\cite{Battistini2015}, who studied main-sequence and sub-giant stars in the local disk. For the heavy elements, there are notable differences in the [La/Fe] and [Ce/Fe] trends between this work and the \cite{Battistini2016} disk sample.} 

Of the neutron-capture elements, we observed rather flat trends in La, Ce and Nd, confirming the high $s$-process contribution to the yields of these elements. The $s$-process element Zr shows a steep decline with metallicity that is comparable to Eu. This observation is in agreement with other studies in the literature, and may also suggest that Zr has a higher $r$-process contribution than expected. [Eu/Fe] decreases with metallicity similarly to the alpha elements. A plateau can be seen at [Fe/H] $\approx -0.5$ dex, after which the Eu abundance ratios decrease rapidly. 

We did not observe consistent latitude-dependent variations in any of the elements. The uniformity of [Na/Fe] with latitude, and the lack of a Na-enhanced population indicate that the double RC observed in the bulge is not due to helium enhancement similar to that observed in globular clusters. Compared the disk/halo samples from GALAH, the over-arching observation is that the bulge follows the same abundance trends as the disk. For Al, which behaves like an alpha element, the bulge is enhanced compared to the low-$\alpha$ disk population. There are also differences between the Galactic components at low metallicity for some elements. In particular, the more metal-poor bulge population ([Fe/H] $\lesssim -0.8$) appears to have enhanced Na, Al, Ni and Cu abundance ratios compared to the disk. {In the previous paper of this series, the metal-poor population also showed enhanced [X/Fe] for some of the alpha elements. Because these elements likely have SNeII origin, our results suggest that this population contains more SNeII material (relative to SNIa) than the more metal-rich bulge population.} For all stars with [Fe/H] $\geq -0.8$, the bulge seems to the share the same chemical evolution as the disk for the light, alpha and iron-peak elements. However, the neutron-capture elements La and Eu indicates that the $r$-process was more dominant in the bulge than in the disk. This result indicates perhaps a higher star formation rate in the bulge compared to the disk. 

Although we confirmed chemical similarities between the disk and bulge for most elements, there are also differences indicating that the stellar population with [Fe/H] $\lesssim -0.8$ has distinct chemistry to the disk/halo at the same metallicity. Furthermore, the neutron-capture elements show that the bulge may have experienced higher star formation rate than the disk. Recently, many studies in the literature have asserted that the metal-rich ([Fe/H] $\gtrsim 0.20$) bulge stars are largely the product of a dynamically buckled inner disk (see e.g., \citealt{DiMatteo2015,Fragkoudi2018}; and \citealt{Nataf2017} for a review). This is supported by the similarity between low-$\alpha$ disk and bulge abundance ratios~\citep[e.g.,][]{Bensby2013,Bensby2017}. However, we have demonstrated that this similarity is not confirmed across the chemical abundance space spanned by our investigation. Differences can be seen not only for neutron-capture elements La and Eu, but also for Al. In addition, Paper I shows that the elements O and Ti follows the high-$\alpha$ disk trend at high metallicity, and remain enhanced compared to the low-$\alpha$ disk. 

These conclusions further highlight the complex evolution in the bulge, which should be investigated further both in terms of alternative models~\cite[e.g.,][]{Inoue2012}; {and with more extensive observational coverage of the inner Galaxy.}

\section*{Acknowledgements}

We thank the anonymous referee for comments that helped to improve this manuscript. LD, MA and KCF acknowledge funding from the Australian Research Council (projects FL110100012 and DP160103747). LD gratefully acknowledges a scholarship from Zonta International District 24. DMN was supported by the Allan C. and Dorothy H. Davis Fellowship. MA's work was conducted as part of the research by Australian Research Council Centre of Excellence for All Sky Astrophysics in 3 Dimensions (ASTRO 3D), through project number CE170100013. Part of this research was conducted at the Munich Institute for Astro- and Particle Physics (MIAPP) of the DFG cluster of excellence ``Origin and Structure of the Universe''. This publication makes use of data products from the Two Micron All Sky Survey, which is a joint project of the University of Massachusetts and the Infrared Processing and Analysis Center/California Institute of Technology, funded by the National Aeronautics and Space Administration and the National Science Foundation. The GALAH survey is based on observations made at the Australian Astronomical Observatory, under programmes A/2013B/13, A/2014A/25, A/2015A/19, A/2017A/18. We acknowledge the traditional owners of the land on which the AAT stands, the Gamilaraay people, and pay our respects to elders past and present. 
\bibliographystyle{mnras}
\bibliography{Ref2} 
\appendix
\section{Detailed spectroscopic analysis}
\subsection{Correlations between effective temperature and abundance ratios}
\label{a2}
\begin{figure*}
	\centering
	\includegraphics[width=1\textwidth]{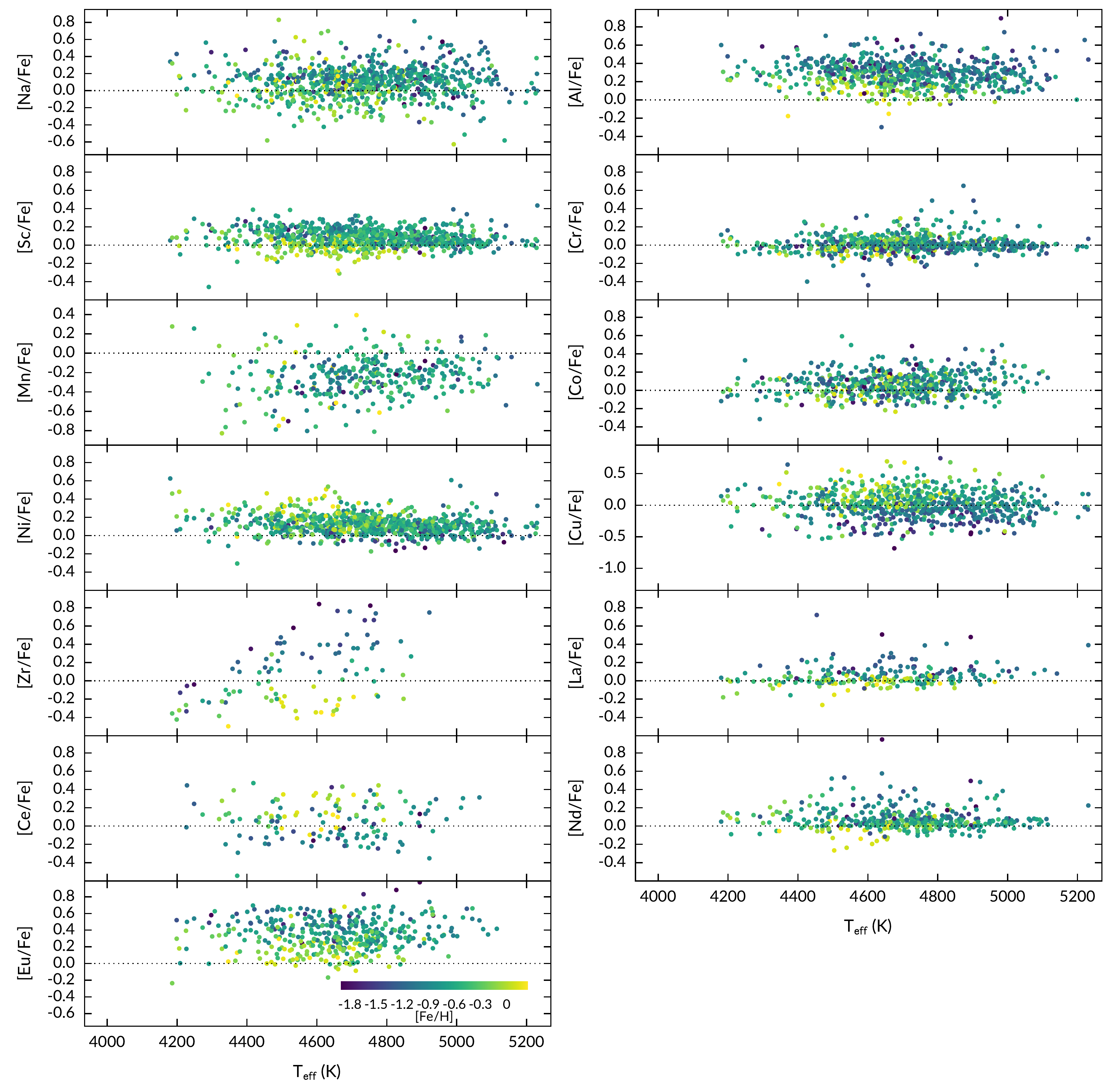}
    \caption{The $T_\mathrm{eff}$-[X/Fe] correlations for all 13 elements. The [Na/Fe] and [Al/Fe] abundance ratios are calculated assuming non-LTE. All other elemental abundances were derived assuming LTE. Most elements do not show a strong correlation with temperature, but positive linear trends can be seen for Mn, Co, although the trend is less significant for Co.}
    \label{fig:abundteff}
\end{figure*}
Fig. \ref{fig:abundteff} shows [X/Fe] as a function of effective temperature (colour-coded by metallicity) for all 13 elements in this study. While most of the abundances were computed in LTE, we do not see significant trends with temperature for the majority of them. [Mn/Fe], however, increases linearly with temperature. Manganese line formation is susceptible to non-LTE effects, which is likely the reason for this abundance-temperature trend~\citep{Bergemann2008,Battistini2015}. The iron-peak element cobalt also shows a very weak
positive linear trend, suggesting departure from LTE~\citep{Bergemann2010}. The non-LTE effect on cobalt is relatively small within our metallicity range, larger effects are observed for metal-poor stars~\citep{Bergemann2010}. Zirconium may have a positive correlation with temperature, but this is difficult to assess as we have few data points, and there is a large range in [Zr/Fe].

{
\subsection{Line analysis for Na, Al and Cu}
In this section we discuss a few lines that are particularly difficult to analyse in cool, metal-rich giants. The \ion{Na}{i} line at 5688 \AA \ and \ion{Cu}{i} line at 5782 \AA \ can be very strong in cool stars. Within our parameter space (T$_\mathrm{eff} \approx$ 4000-5000 K; $\log g \approx$ 3.5--1.5 cms$^{-2}$), these line profiles can be measured fairly reliably (our results are . The metal-rich ([Fe/H] $>$ 0) stars in our sample have higher temperature and $\log g$ and our coolest stars are also among the most metal poor, which means that neither line becomes saturated as shown in Fig. \ref{fig:na} and \ref{fig:cu}. However, due to the strength of these lines, the [Na/Fe] and [Cu/Fe] abundance ratios may be sensitive to micro-turbulence. 

\begin{figure}
	\centering
	\includegraphics[width=1\columnwidth]{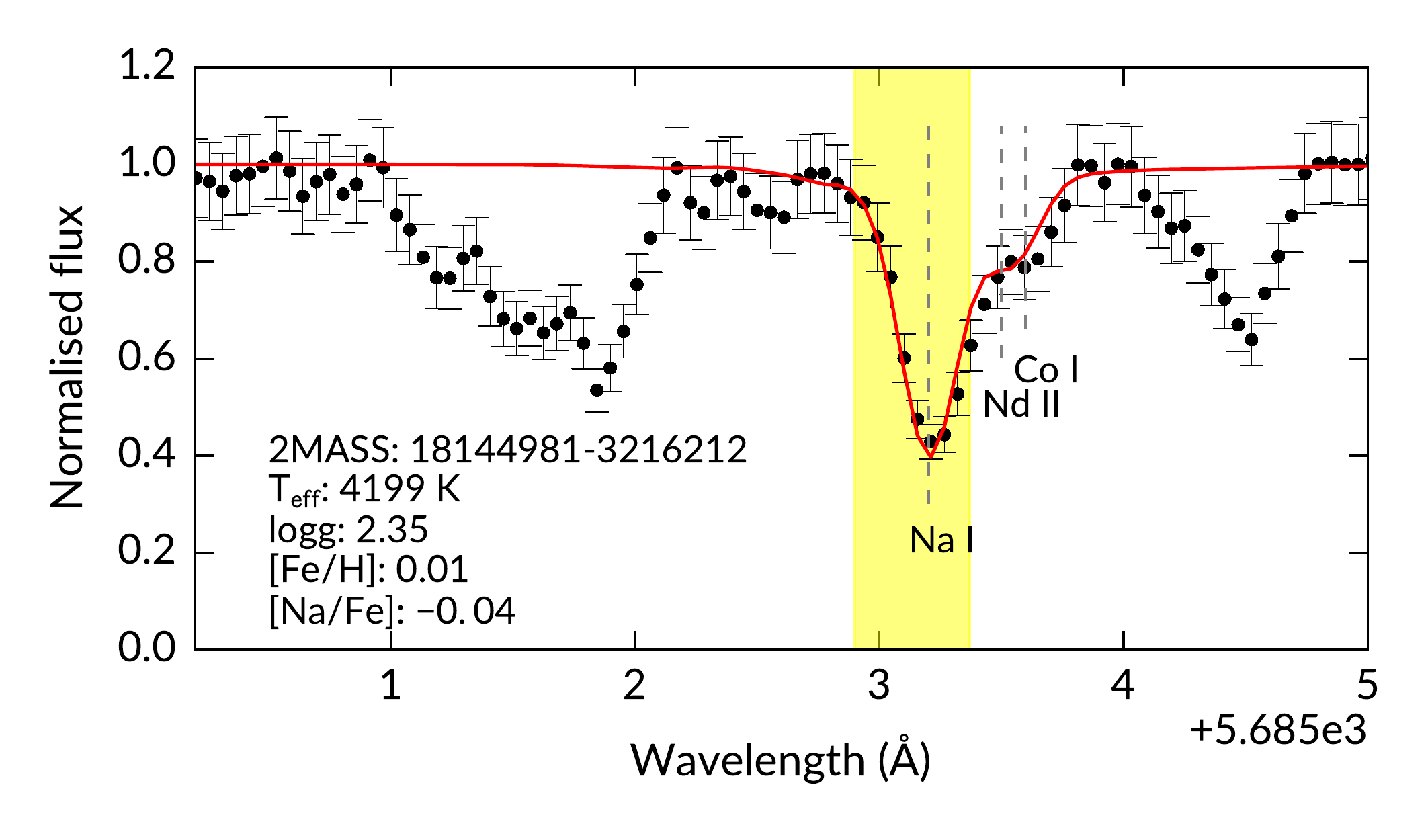}
    \caption{Line synthesis of \ion{Na}{i}. Black points are the normalised observed flux and the red line is the final model. The shaded part indicates the line mask - i.e. pixels used to determine [Na/Fe]. The blending lines are excluded from the line mask.}
    \label{fig:na}
\end{figure}

Although \ion{Na}{i} is blended with \ion{Nd}{ii} and \ion{Co}{i}, the blending is not too severe and is accounted for. Fig. \ref{fig:na} shows the Na line synthesis for a typical star in the sample. The shaded part indicates the pixels used for abundance optimisation: blending lines \ion{Nd}{ii} and \ion{Co}{i} are excluded.

The \ion{Cu}{i} line profile may be affected by an intersellar band at 5780 \AA \ (which is not included in the Cu line segment in Fig \ref{fig:cu}). However when inspecting the sample, we did not find cases where the \ion{Cu}{i} line profile is peculiar. Where the line profile is severely affected, the optimisation routine will fail converge and will not provide a final abundance ratio.

\begin{figure}
	\centering
	\includegraphics[width=1\columnwidth]{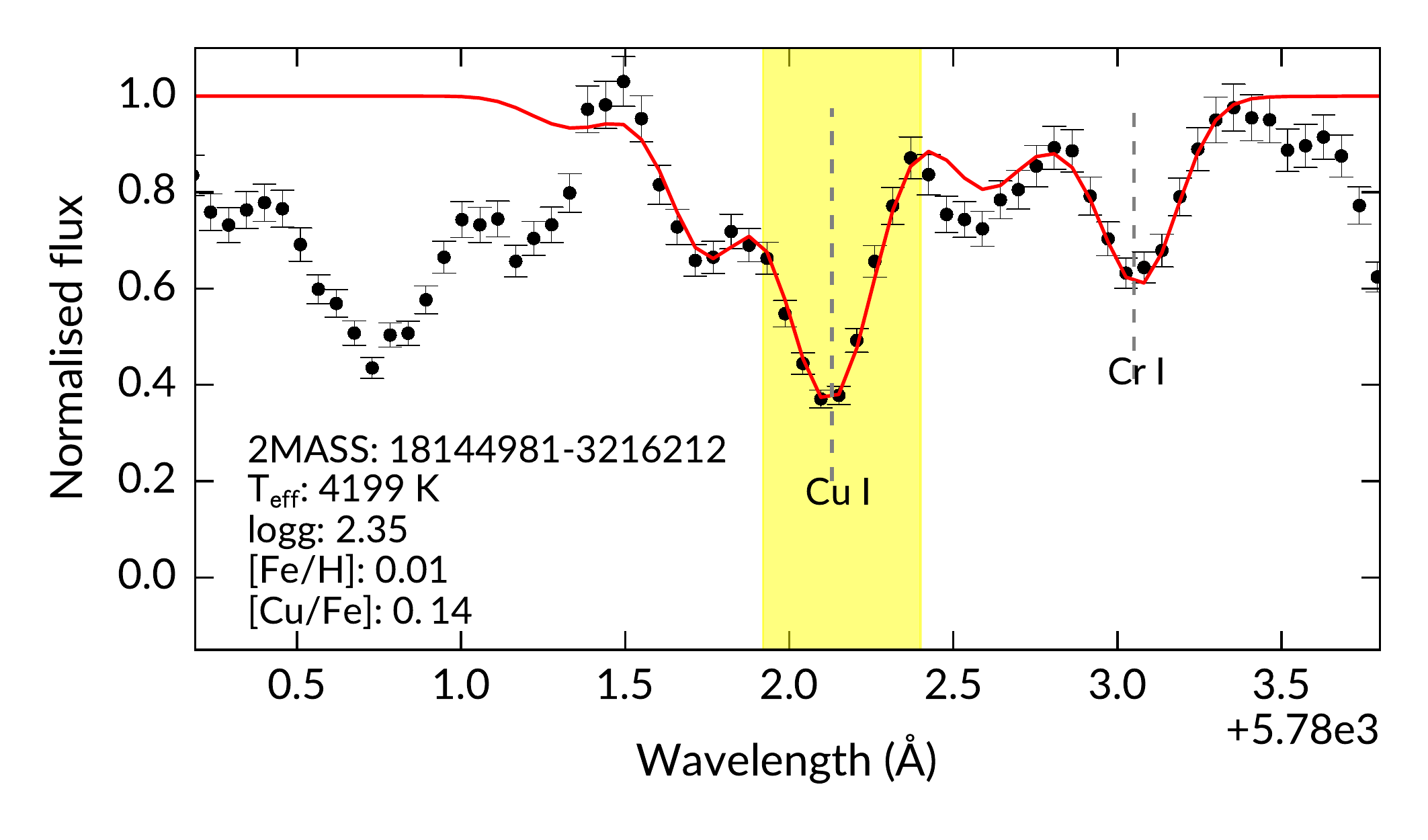}
    \caption{Line synthesis of \ion{Cu}{i}. Black points are the normalised observed flux and the red line is the final model. The shaded part indicates the line mask - i.e. pixels used to determine [Cu/Fe].}
    \label{fig:cu}
\end{figure}

\begin{figure}
	\centering
	\includegraphics[width=1\columnwidth]{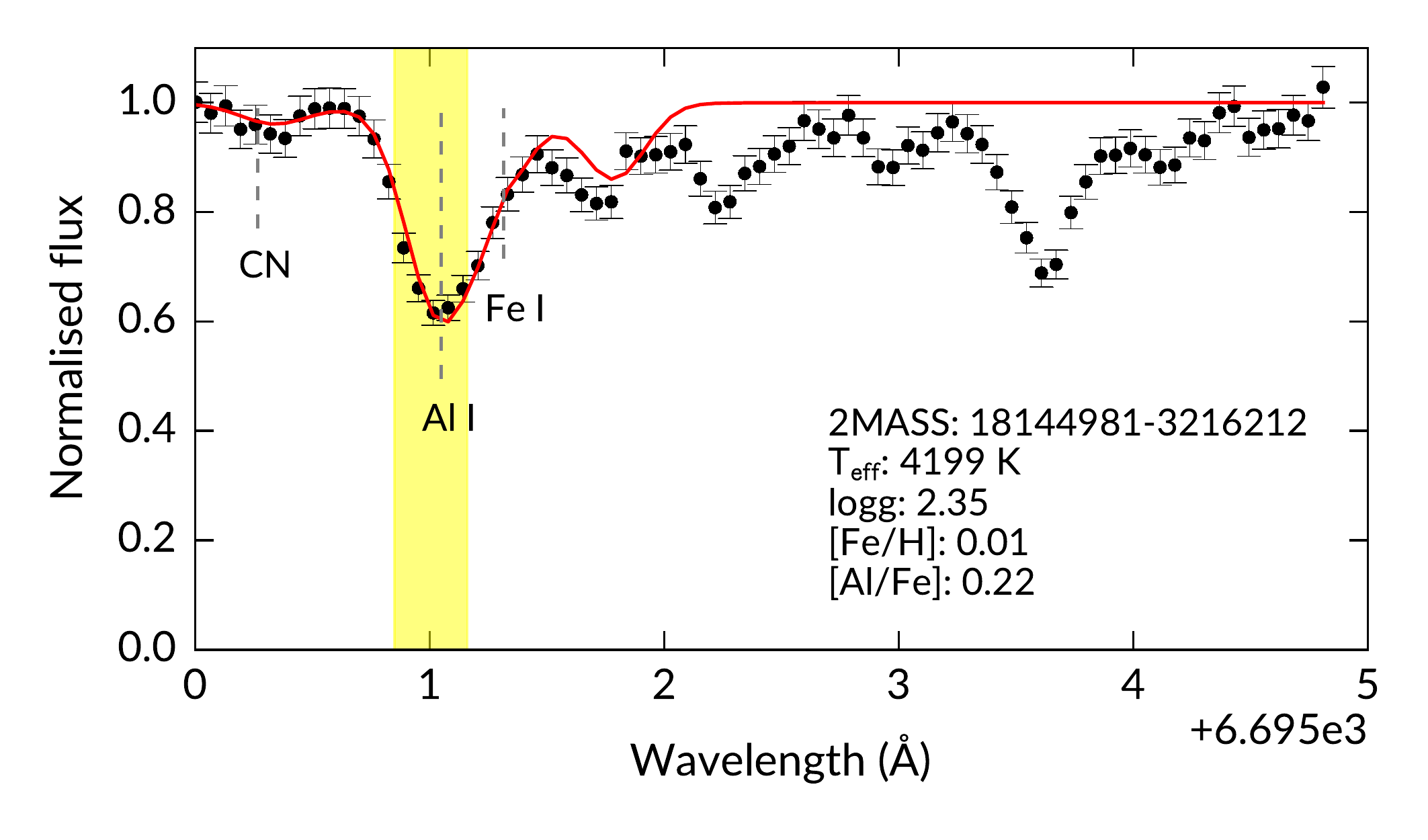}\\
	\includegraphics[width=1\columnwidth]{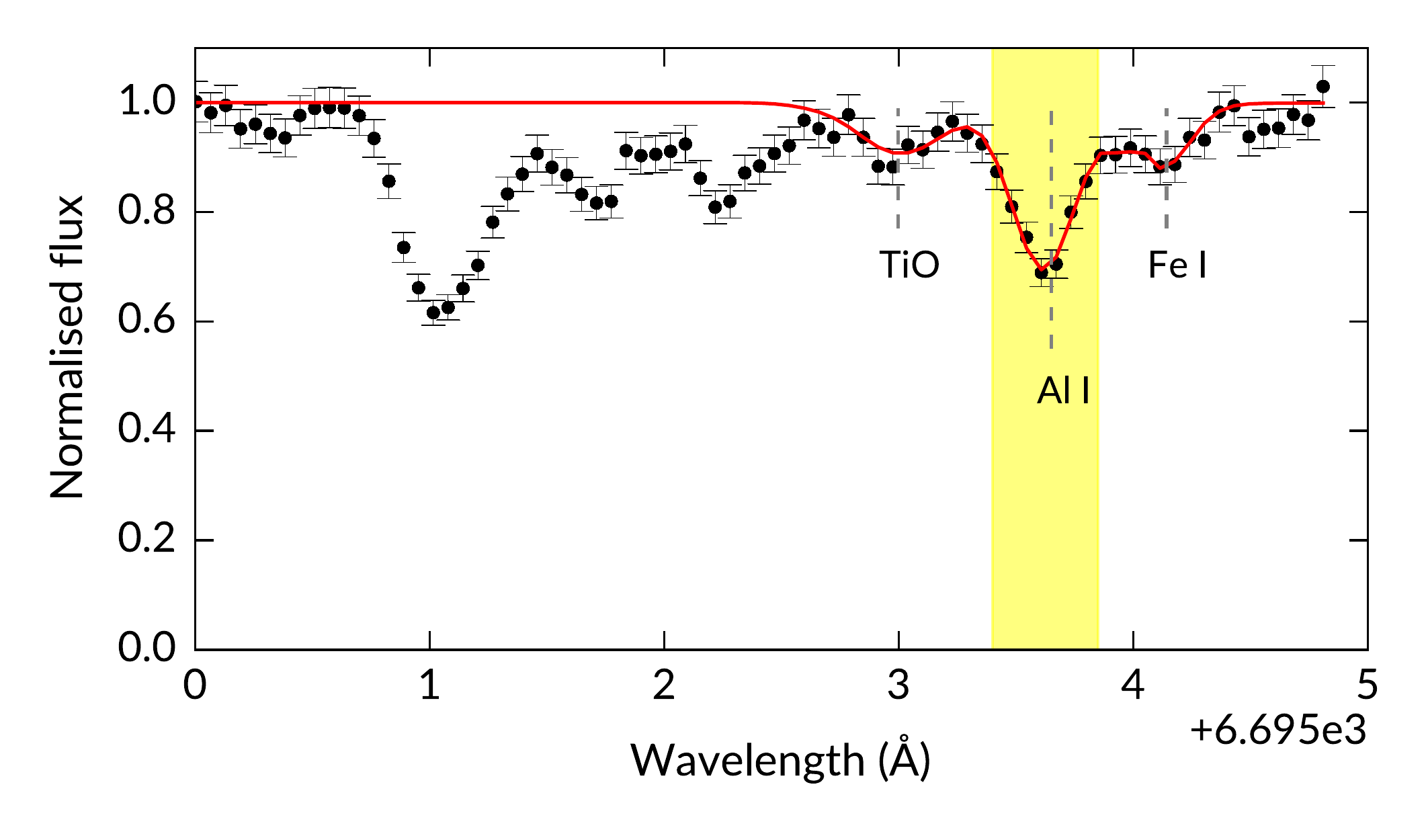}
    \caption{Line synthesis of \ion{Al}{i} at 6696 and 6699 \AA. Black points are the normalised observed flux and the red line is the final model. The shaded part indicates the line mask - i.e. pixels used to determine [Al/Fe]. For \ion{Al}{i} 6696, the blending \ion{Fe}{i} line is excluded from the line mask. Molecular lines are also included in the line synthesis.}
    \label{fig:al}
\end{figure}

The 6696/6698 \ion{Al}{i} lines can also be very strong in cool, metal-rich giants, but we found that they are of suitable strength for our sample. Fig \ref{fig:al} also shows that the neighbouring molecular lines are included in the synthesis of Al lines, and for \ion{Al}{i} 6696, the blending \ion{Fe}{i} is excluded from the line mask. 

\subsection{Arcturus abundance ratios}
To further test the accuracy of our method, we compare our abundance measurements for the benchmark giant Arcturus with literature studies in Table \ref{arcturus}. We measured abundance ratios for the HERMES observation of Arcturus using stellar parameters derived in Paper I ($T_\mathrm{eff}$ = 4381, $\log g$ = 1.66, [Fe/H] = $-0.57$). The comparison shows that for most elements, there are very small ($<0.05$ dex) or no systematic offset between our and literature results when uncertainties are taken into account. However a number of elements indicate $\lesssim$0.1 dex offset, even when uncertainties are taken into account, these are: Na, Mn, Zr and Nd. 

\begin{table*}
\caption{Abundance ratios [X/Fe] for Arcturus from this work and recent literature studies. Our abundance ratios were measured for the HERMES observation of Arcturus. Literature studies provided abundance ratios for the Arcturus atlas~\protect\citep{Hinkle2000}; \protect\cite{Jofre2015} also included a spectrum observed with the NARVAL spectrograph.}
\centering
\begin{tabular}{clccc}
\hline
Element & This work & \cite{Ramirez2011} & \cite{Jofre2015} & \cite{Johnson2012,Johnson2014}\\
\hline
Na & +0.22 $\pm$ 0.01 & +0.11 $\pm$ 0.03 & - & +0.06\\
Al & +0.35 $\pm$ 0.004 & +0.34 $\pm$ 0.03  & - & +0.31\\
Sc & +0.15 $\pm$ 0.004 & +0.15 $\pm$ 0.08 & +0.09 $\pm$ 0.06 & - \\
Cr & $-0.04$ $\pm$ 0.005 & $-0.05 \ \pm $ 0.04 & $-0.06$ $\pm$ 0.04 & $-0.08$ \\
Mn & $-0.18$ $\pm$ 0.02 & - & $-0.37$ $\pm$ 0.09 & - \\
Co & +0.11 $\pm$ 0.007 & +0.09 $\pm$ 0.04 & +0.11 $\pm$ 0.05  & +0.12\\
Ni & +0.11 $\pm$ 0.005 & +0.06 $\pm$ 0.03 & +0.03 $\pm$ 0.05 & +0.06 \\
Cu & +0.15 $\pm$ 0.01 & - & - & +0.17 \\
Zr & $-0.12$ $\pm$ 0.014 & - & - & +0.00\\
La & $-0.09$ $\pm$ 0.014 & - & - & $-0.06$ \\
Ce & $-0.18$ $\pm$ 0.02 & - & - & -\\
Nd & $-0.04$ $\pm$ 0.012 & - & - & +0.05 \\
Eu & +0.24 $\pm$ 0.01 & - & - & +0.29\\
\hline
\end{tabular}
\label{arcturus}
\end{table*}

\subsection{Comparison of HERBS and GALAH DR2 abundances}
\label{herbs-galah-comp}
The HERBS and GALAH abundance analyses utilise the same spectral synthesis code (and version), normalisation method, convergence criteria and atomic data to ensure that the disk and bulge results are comparable. However, in some cases we have omitted unsuitable lines from the GALAH linelist, and synthesised individual lines before taking the weighted average as the final abundance ratio. 
To assess if there are any offsets that may affect our conclusions in Section \ref{sec4}, we analysed 100 giants from the GALAH comparison sample with our pipeline and compared the results with GALAH DR2. The differences are given in Table \ref{table:herbs-galah-comp} along with typical uncertainties for each element (HERBS and GALAH reported similar uncertainties). 

The elements that show [X/Fe] offsets that are significant compared to the typical uncertainties are Na, Al, Sc and Cu. We comment on these elements below.
\begin{itemize}
    \item Our [Na/Fe] values are underestimated compared to GALAH DR2. We observed that the bulge has higher [Na/Fe] than the disk at [Fe/H] $< -0.8$, so this offset means that the difference could be even more pronounced. 
    \item We also observed that the bulge seems to be enhanced in Al compared to the disk at [Fe/H] $< -0.8$. The bulge median is this metallicity is $0.34 \pm 0.05$; the disk median is $0.19 \pm 0.09$ (but note that the GALAH disk sample has some very high [Al/Fe] at low metallicity). The offset in Al is smaller than the observed difference, but reduces its significance.
    \item For Sc, using the lines we selected indeed lowered [Sc/Fe] ratios of the GALAH sample by $\approx$0.2 dex. Due to the large systematic difference, we did not compare GALAH and HERBS results. 
    \item For Cu, we did not observe differences when the full bulge and disk data sets are compared; but the difference indicates that our results are underestimated and therefore the bulge could have slightly higher [Cu/Fe] than the disk, especially at [Fe/H] $< -0.8$.
\end{itemize}

\begin{table}
\caption{Mean differences between abundance ratios obtained with the HERBS and GALAH analysis. The mean differences (and standard deviations) were computed for elements that HERBS and GALAH have in common: Na, Al, Sc, Cr, Co, Ni, Cu, La and Eu (see Fig. \ref{fig:galah-comp}). The typical uncertainty for each element is given as $\sigma_\mathrm{[X/Fe]}$.}
\label{table:herbs-galah-comp}
\begin{tabular}{ccc}
\hline
Element & $\Delta$[X/Fe] (HERBS $-$ GALAH) & Typical $\sigma_\mathrm{[X/Fe]}$ \\
\hline
Na & $-0.11$ $\pm$ 0.05 & 0.03 \\
Al & +0.08 $\pm$ 0.09 & 0.03 \\
Sc & $-0.21$ $\pm$ 0.07 & 0.03 \\
Cr & $-0.05$ $\pm$ 0.08 & 0.04 \\
Co & +0.02 $\pm$ 0.04 &  0.03 \\
Ni & $-0.05$ $\pm$ 0.07 &  0.04 \\
Cu & $-0.12$ $\pm$ 0.03 & 0.04 \\
La & $-0.00$ $\pm$ 0.03 & 0.05 \\
Eu & $-0.02$ $\pm$ 0.03 & 0.03 \\
\hline
\end{tabular}
\end{table}
}
\section{Data tables}
\subsection{Linelist}
\label{a1}
Table \ref{table:atom} contains a list of lines used to derive the abundance ratios in this work. While the atomic data is the same as that of GALAH Data Release 2~\citep{Buder2018}, we did not necessarily use every line in the GALAH linelist for each element.
\begin{table}
\label{table:atom}
\caption{The full list of lines and atomic data used to derive abundance ratios in this work.}
\centering
\begin{tabular}{cccc}
\hline
Species & Wavelength (\AA) & $\log \left(gf\right)$ & Excitation potential (eV)\\
\hline
\ion{Na}{i} & 5688.2050 & -0.404 & 2.104\\
\ion{Al}{i} & 6696.0230 & -1.569 & 3.143\\
\ion{Al}{i} & 6698.6730 & -1.870 & 3.143\\
\ion{Sc}{i} & 4753.1610 & -1.659 & 0.000\\
\ion{Sc}{i} & 5671.8163 & -0.290 & 1.448\\
\ion{Sc}{i} & 5686.8386 & -0.133 & 1.440\\
\ion{Sc}{ii} & 6604.6010 & -1.309 & 1.357\\
\ion{Cr}{i} & 5719.8150 & -1.580 & 3.013\\
\ion{Cr}{i} & 5787.9190 & -0.083 & 3.322\\
\ion{Cr}{i} & 5844.5950 & -1.770 & 3.013\\
\ion{Mn}{i} & 4761.5060 & -0.548 & 2.953\\
\ion{Mn}{i} & 4765.8525 & -0.445 & 2.941\\
\ion{Co}{i} & 6632.4505 & -2.726 & 2.280\\
\ion{Ni}{i} & 5748.3507 & -3.240 & 1.676\\
\ion{Ni}{i} & 5846.9935 & -3.460 & 1.676\\
\ion{Ni}{i} & 6482.7983 & -2.630 & 1.935\\
\ion{Ni}{i} & 6532.8730 & -3.350 & 1.935\\
\ion{Ni}{i} & 6586.3098 & -2.780 & 1.951\\
\ion{Ni}{i} & 6643.6303 & -2.220 & 1.676\\
\ion{Cu}{i} & 5782.1554 & -1.789 & 1.642\\
\ion{Zr}{i} & 4805.8700 & -0.420 & 0.687\\
\ion{Zr}{i} & 4828.0400 & -0.640 & 0.623\\
\ion{La}{ii} & 4716.4400 & -1.210 & 0.772\\
\ion{La}{ii} & 4748.7300 & -0.540 & 0.927\\
\ion{La}{ii} & 4804.0690 & -1.490 & 0.235\\
\ion{La}{ii} & 5805.7700 & -1.560 & 0.126\\
\ion{Ce}{ii} & 4773.9410 & -0.390 & 0.924\\
\ion{Nd}{ii} & 4811.3420 & -1.140 & 0.064\\
\ion{Nd}{ii} & 5740.8580 & -0.530 & 1.160\\
\ion{Nd}{ii} & 5811.5700 & -0.860 & 0.859\\
\ion{Eu}{ii} & 6645.0978 & +0.120 & 1.380\\
\hline
\end{tabular}
\end{table}
\subsection{Abundances catalogue}
\label{a3}
The full data catalogue accompanying this paper is provided as supporting material, accessible through the publisher website. The contents of the catalogue is described in Table~\ref{table:data}.

\begin{table}
	\caption{Description of the data catalogue. The uncertainties in abundance ratios are $\chi^2$ fitting errors.}
	\label{table:data}
	\begin{tabular}{lll}
		\hline 
        Column & Name & Description\\
        \hline
        {[1]} & 2MASS ID & The 2MASS identifier of the star\\
        {[2]} & RAJ2000 & The right ascension at epoch J2000 (degrees)\\
        {[3]} & DECJ2000 & The declination at epoch J2000 (degrees)\\
        {[4]} & $T_\mathrm{eff}$ & Effective temperature (K)\\
        {[5]} & $\sigma_\mathrm{T_{eff}}$ & Uncertainty in effective temperature (K)\\
        {[6]} & $\log g$ & Surface gravity (cm s$^{-2}$)\\
        {[7]} & $\sigma_{\log g}$ & Uncertainty in surface gravity (cm s$^{-2}$)\\
        {[8]} & [Fe/H] & Metallicity\\
		{[9]} & $\sigma_\mathrm{[Fe/H]}$ & Uncertainty in metallicity\\
        {[10]} & [Na/Fe] & Abundance ratio for Na\\
        {[11]} & $\sigma_\mathrm{[O/Fe]}$ & Uncertainty in [Na/Fe]\\
        {[12]} & [X/Fe] & Same as [10], but for Al\\
        {[13]} & $\sigma_\mathrm{[X/Fe]}$ & Same as [11], but for Al\\
        {[14]} & [X/Fe] & Same as [10], but for Sc\\
        {[15]} & $\sigma_\mathrm{[X/Fe]}$ & Same as [11], but for Sc\\
        {[16]} & [X/Fe] & Same as [10], but for Cr\\
        {[17]} & $\sigma_\mathrm{[X/Fe]}$ & Same as [11], but for Cr\\
        {[18]} & [X/Fe] & Same as [10], but for Mn\\
        {[19]} & $\sigma_\mathrm{[X/Fe]}$ & Same as [11], but for Mn\\
        {[20]} & [X/Fe] & Same as [10], but for Co\\
        {[21]} & $\sigma_\mathrm{[X/Fe]}$ & Same as [11], but for Co\\
        {[22]} & [X/Fe] & Same as [10], but for Ni\\
        {[23]} & $\sigma_\mathrm{[X/Fe]}$ & Same as [11], but for Ni\\
        {[24]} & [X/Fe] & Same as [10], but for Cu\\
        {[25]} & $\sigma_\mathrm{[X/Fe]}$ & Same as [11], but for Cu\\
        {[26]} & [X/Fe] & Same as [10], but for Zr\\
        {[27]} & $\sigma_\mathrm{[X/Fe]}$ & Same as [11], but for Zr\\
        {[28]} & [X/Fe] & Same as [10], but for La\\
        {[29]} & $\sigma_\mathrm{[X/Fe]}$ & Same as [11], but for La\\
        {[30]} & [X/Fe] & Same as [10], but for Ce\\
        {[31]} & $\sigma_\mathrm{[X/Fe]}$ & Same as [11], but for Ce\\
        {[32]} & [X/Fe] & Same as [10], but for Nd\\
        {[33]} & $\sigma_\mathrm{[X/Fe]}$ & Same as [11], but for Nd\\
        {[34]} & [X/Fe] & Same as [10], but for Eu\\
        {[35]} & $\sigma_\mathrm{[X/Fe]}$ & Same as [11], but for Eu\\
        \hline
	\end{tabular}
\end{table}
\bsp	
\label{lastpage}
\end{document}